\documentclass[aps,pra,amsmath,amssymb,floatfix,twocolumn,showpacs,10pt,nofootinbib]{revtex4-1}
 
\usepackage{times}
\usepackage{bbold}
\usepackage{bm}
\usepackage{graphicx} 
\usepackage{color}
\usepackage{hyperref}
\usepackage{textcomp}
\usepackage{amssymb} 
\usepackage{wasysym}
\usepackage{mathptmx}

\begin{document}

\title{Effects of nuclear spins on the transport properties of the edge of two-dimensional topological insulators}

\author{Chen-Hsuan Hsu$^{1}$}
\author{Peter Stano$^{1,2,3}$}
\author{Jelena Klinovaja$^{1,4}$}
\author{Daniel Loss$^{1,4}$}

\affiliation{$^{1}$RIKEN Center for Emergent Matter Science (CEMS), Wako, Saitama 351-0198, Japan}
\affiliation{$^{2}$Department of Applied Physics, School of Engineering, University of Tokyo, 7-3-1 Hongo, Bunkyo-ku, Tokyo 113-8656, Japan}
\affiliation{$^{3}$Institute of Physics, Slovak Academy of Sciences, 845 11 Bratislava, Slovakia}
\affiliation{$^{4}$Department of Physics, University of Basel, Klingelbergstrasse 82, CH-4056 Basel, Switzerland}

\date{\today}

\begin{abstract}
The electrons in the edge channels of two-dimensional topological insulators can be described as a helical Tomonaga-Luttinger liquid. They couple to nuclear spins embedded in the host materials through the hyperfine interaction, and are therefore subject to elastic spin-flip backscattering on the nuclear spins. We investigate the nuclear-spin-induced edge resistance due to such backscattering by performing a renormalization-group analysis. Remarkably, the effect of this backscattering mechanism is stronger in a helical edge than in nonhelical channels, which are believed to be present in the trivial regime of InAs/GaSb quantum wells. In a system with sufficiently long edges, the disordered nuclear spins lead to an edge resistance which grows exponentially upon lowering the temperature. On the other hand, electrons from the edge states mediate an anisotropic Ruderman-Kittel-Kasuya-Yosida nuclear spin-spin interaction, which induces a spiral nuclear spin order below the transition temperature. We discuss the features of the spiral order, as well as its experimental signatures. In the ordered phase, we identify two backscattering mechanisms, due to charge impurities and magnons. The backscattering on charge impurities is allowed by the internally generated magnetic field, and leads to an Anderson-type localization of the edge states. The magnon-mediated backscattering results in a power-law resistance, which is suppressed at zero temperature. Overall, we find that in a sufficiently long edge the nuclear spins, whether ordered or not, suppress the edge conductance to zero as the temperature approaches zero. 
\end{abstract}
 
\maketitle

\section{Introduction}

The helical edge states are the essential feature of two-dimensional topological insulators (2DTIs), whose examples include HgTe/(Hg,Cd)Te quantum wells~\cite{Bernevig:2006,Konig:2007,Roth:2009,Gusev:2013,Gusev:2014,Ma:2015,Olshanetsky:2015,Deacon:2017}
and InAs/GaSb quantum wells~\cite{Liu:2008,Knez:2011,Suzuki:2013,Charpentier:2013,Knez:2014,Suzuki:2015,Mueller:2015,Qu:2015,Li:2015,Du:2015,Nichele:2016,Couedo:2016,Akiho:2016,Nguyen:2016,Mueller:2017}.
Since elastic backscattering of the helical edge states must be accompanied by spin flips, the edge channel conductance is insensitive to perturbations that respect time-reversal symmetry. However, it remains to be clarified whether alternative backscattering mechanisms, arising from the broken time-reversal symmetry or inelastic processes, can still cause a substantial resistance to destroy the edge conductance quantization. Among other proposed mechanisms~\cite{Wu:2006,Xu:2006,Maciejko:2009,Jiang:2009,Strom:2010,Tanaka:2011,Hattori:2011,Lunde:2012,Budich:2012,Lezmy:2012,Schmidt:2012,Delplace:2012,Crepin:2012,DelMaestro:2013,Altshuler:2013,Vayrynen:2013,Geissler:2014,Kainaris:2014,Vayrynen:2014,Chou:2015,Yevtushenko:2015,Vayrynen:2016,Xie:2016,Kharitonov:2017,Wang:2017,Chou:2017,Groenendijk:2017,Vayrynen:2018}, 
in Ref.~\cite{Hsu:2017} we pointed out that in general nuclear spins could be such a resistance source of 2DTIs.

Nuclear spins, typically present in usual 2DTI host materials, couple to the electrons in the edge channels via the hyperfine interaction, and thus result in elastic spin-flip backscattering. It is therefore necessary to examine whether the nuclear spins are detrimental to the edge states. Furthermore, because the edge states in InAs/GaSb heterostructure persist even in the regime where the energy band is not inverted~\cite{Nichele:2016,Nguyen:2016}, the topological (helical) character of the edge states in the band-inverted (nominally topological) regime remains to be verified. 

This work extends results obtained in Ref.~\cite{Hsu:2017}, the main findings of which are summarized as follows. 
First, as in non-topological systems~\cite{Giamarchi:1988,Giamarchi:2003}, the electron-electron interaction strongly enhances the backscattering effects in one dimension. Therefore, in strongly interacting systems the effects of the nuclear spins on edge resistance become prominent in spite of the typically weak hyperfine couplings in 2DTI host materials~\cite{Gueron:1964,Schliemann:2003,Braun:2006,Lunde:2013}. 
Second, the resistance caused by randomly oriented nuclear spins is bigger for a longer edge, a lower temperature, and stronger electron-electron interactions. Third, a nuclear spin order is stabilized by the Ruderman-Kittel-Kasuya-Yosida (RKKY) interaction. Finally, the transport properties of the edge states are influenced by the ordering of the nuclear spins. The impurities\footnote{To avoid possible confusion, we will use the term ``impurity'' when we refer to charge (nonmagnetic) impurity, and reserve the term ``(dis-)order'' for the (dis-)ordering of the nuclear spins.} and magnons enable additional backscattering processes in the ordered phase. The resistance due to the impurities ultimately dominates over the one from the magnons as the temperature approaches zero, leading to our conclusion that in general the nuclear spins suppress the edge conductance at very low temperatures.
 
The topic is, however, rather complex due to the intertwining of electron and nuclear subsystems. Therefore, a separate account is necessary. Specifically, here we additionally provide the details of our renormalization-group (RG) analysis on the nuclear-spin-induced backscattering in the bosonization framework, giving the reader a comprehensive understanding of various backscattering mechanisms in the disordered and ordered phases.  
We also compare the effects of the nuclear spins in various materials, including InAs/GaSb and HgTe/(Hg,Cd)Te quantum wells, and GaAs quantum wires. Remarkably, this provides signatures to reveal the helical nature of the edge states, since disordered nuclear spins lead to a stronger effect in a helical edge in InAs/GaSb than in nonhelical channels in a GaAs wire or in the trivial regime of InAs/GaSb, despite comparable values of material parameters. 
Further, we investigate the RKKY-induced nuclear spin order, whose elaborate description was not included in our previous work. Here we show how the helical edge states mediate an anisotropic RKKY interaction,\footnote{Whereas here we focus on the RKKY interaction mediated by the edge states of a 2DTI, we note that the behavior of the RKKY interaction on the surfaces of three-dimensional topological insulators can also be very rich~\cite{Zyuzin:2014b,Klier:2017}.} inducing a spiral nuclear spin order.\footnote{The term ``spiral nuclear spin order'' is used to distinguish the order in a 2DTI edge from the nuclear spin helix in a nonhelical channel, as will be explained in Sec.~\ref{SubSec:RKKY}. Moreover, we adopt this term also in order to avoid possible confusion between the ordering of the nuclear spins and the helicity of the edge states.} We also explain in detail the differences between the order in a helical edge and the one in a spin-degenerate, nonhelical wire. Remarkably, the helical character of the edge states enhances the instability toward the nuclear spin ordering. In addition, we discuss experimental signatures of the spiral nuclear spin order in optical and charge transport measurements, which gives guidance for  possible experimental verifications of the formation of the spiral order predicted in our work.

The additional backscattering mechanisms due to the spiral order also deserves further elucidation. 
Since a macroscopic magnetic (Overhauser) field originating from the nuclear spin order admixes the edge states with the opposite spins, the edge states do not retain perfect helicity anymore, becoming susceptible to the backscattering on impurities~\cite{Hsu:2017}. Here we present the derivation of the Overhauser-field-assisted backscattering action, and show that the form can be understood from the spin rotational symmetry of electron-electron interaction. 
Finally, here we give the derivation of the effective action due to the magnon-mediated backscattering, including the emission and the absorption processes. Whereas the efficiency of these processes depends on the magnon energy and the magnon occupation, thus leading to rich temperature dependence of the resulting edge resistance, 
the derived effective action allows us to intuitively visualize the effects of the magnon-mediated backscattering on the resistance.
  
This paper is organized as follows. In Sec.~\ref{Sec:Hamiltonian}, we introduce the Hamiltonian, including the edge-state electron Hamiltonian and the hyperfine interaction between the electron spins and the nuclear spins. 
In Sec.~\ref{Sec:hf-bs}, we investigate the backscattering caused by disordered nuclear spins and compare the localization effect in various materials and experimental setups. Using an RG analysis we calculate the edge resistance in the short-edge, the high-temperature, and the strong-coupling regimes, as well as the differential resistance in the high-bias regime. 
In Sec.~\ref{Sec:ordering}, we investigate the RKKY-induced spiral nuclear spin order. We first compare the spiral order in a 2DTI with the nuclear spin helix in a spin-degenerate wire in Sec.~\ref{SubSec:RKKY} and then discuss its experimental signatures in Sec.~\ref{SubSec:signature}. 
In addition, we examine the self-consistency condition of the RKKY approach in Sec.~\ref{SubSec:consistency}. 
In Sec.~\ref{Sec:ordered}, we examine the effects of the magnons and the impurities on the 2DTI edge states in the ordered phase. In Sec.~\ref{SubSec:hx-bs} we analyze the resistance due to the Overhauser-field-assisted backscattering on impurities, and find that it gives rise to a resistance that is comparable to the one due to the disordered nuclear spins.
We then discuss the features to distinguish the two resistance sources.
In Sec.~\ref{SubSec:mag-bs}, we calculate the resistance due to the magnon-mediated backscattering, including both the emission and the absorption processes. 
Finally, in Sec.~\ref{Sec:discussion} we summarize the resistance from all these backscattering mechanisms
and discuss the relevance of nuclear spins to the observed edge resistance in experiments.
The technical details are presented in the appendices.

\section{Hamiltonian~\label{Sec:Hamiltonian}} 

\begin{figure}[t]
\centering
\includegraphics[width=\linewidth]{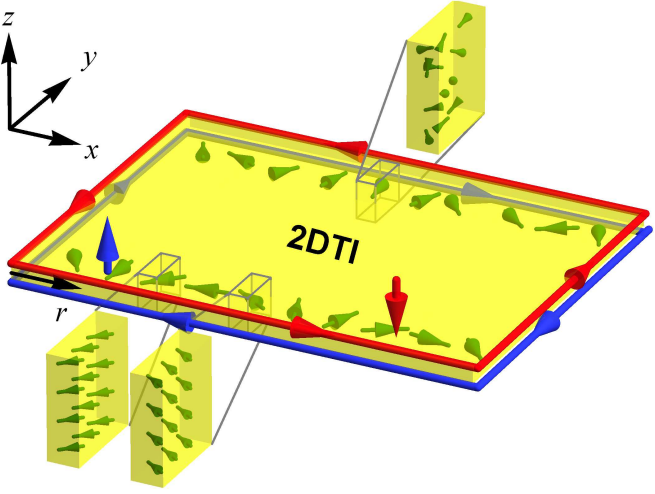}
\caption{In a 2DTI of rectangular shape, electrons propagate along the edges. In this work, without loss of generality, we focus on one of the four edges (labeled by the coordinate $r$), where the up-spin electron $L_{\uparrow}$ (blue) moves in the left direction, and the down-spin electron $R_{\downarrow}$ (red) moves in the right direction. 
The electron spin quantization axis is defined as the $z$ axis, which is perpendicular to the 2DTI plane. Nuclear spins (indicated by the green arrows) are randomly oriented (displayed at the top edge) above the transition temperature $T_{0}$, and form a spiral order below $T_{0}$. In the ordered phase, the nuclear spins align ferromagnetically within each cross section, and rotate along the edge, as demonstrated at the bottom edge.
For clarity, the spins are drawn only for two opposite  edges, and the spin up and spin down edge states are separated spatially.}
\label{Fig:setup}
\end{figure}

We begin by modeling the edge states and the nuclear spins with the Hamiltonian, $\mathit{H}=\mathit{H}_{\textrm{el}}+\mathit{H}_{\textrm{hf}}$. Throughout the paper we assume that the edge states consist of the right-moving down-spin $R_{\downarrow}$ and the left-moving up-spin $L_{\uparrow}$ electrons (see Fig.~\ref{Fig:setup}; the opposite edge of the 2DTI is assumed to be far away, and decoupled from the edge under consideration). In the bosonization framework, we describe these edge states as a helical Tomonaga-Luttinger liquid~\cite{Wu:2006,Xu:2006},
\begin{equation}
\mathit{H}_{\textrm{el}} = \int \frac{\hbar dr}{2\pi} \, \left\{ uK \left[ \partial_{r} \theta(r) \right]^2 + \frac{u}{K} \left[ \partial_{r} \phi(r) \right]^2 \right\},
\label{Eq:H_el}
\end{equation}
where the bosonic fields $\theta$ and $\phi$ relate to the original $R_{\downarrow}$ and $L_{\uparrow}$ fermionic fields through 
\begin{subequations}
\label{Eq:bosonization}
\begin{eqnarray}
R_{\downarrow} (r) &=& \frac{U_{R}}{\sqrt{2\pi a}} e^{i k_{F}r} e^{i[-\phi(r) + \theta(r)]}, \\
L_{\uparrow} (r) &=& \frac{U_{L}}{\sqrt{2\pi a}} e^{-i k_{F}r} e^{i[\phi(r) + \theta(r)]},
\end{eqnarray}
\end{subequations} 
with the Klein factors $U_{R}$ and $U_{L}$, the Fermi wave number $k_F$, and the edge coordinate $r$. The electron-electron interaction is parametrized by the Luttinger liquid parameter $K$. Here $u=v_{F}/K$ is the renormalized velocity with the Fermi velocity $v_{F}$. 
The short-distance cutoff $a$, required by the bosonization prescription, is taken as the transverse decay length of the wave function of the edge states, $a=\hbar v_{F}/\Delta$, with the 2DTI bulk gap $\Delta$.\footnote{Since the energy bands of the edge states merge into the bulk energy bands above the 2DTI gap $\Delta$, the helical Tomonaga-Luttinger liquid description of the edge states breaks down above $\Delta$ (here the energies are measured from the Dirac point). Therefore, for our bosonization procedure we choose $\Delta$ as the high-energy cutoff, below which the edge states retain their linear energy dispersion and helical nature. The corresponding short-distance cutoff is thus given by the evanescent decay length of the edge-state electron wave function into the bulk~\cite{Maciejko:2009,Kharitonov:2017}.} 
The Fermi energy is given by $\epsilon_{F} \equiv \hbar v_{F} k_{F}/2$. The helical Tomonaga-Luttinger liquid action can be obtained by integrating out the $\theta$ field in Eq.~(\ref{Eq:H_el}),   
\begin{eqnarray}
\frac{\mathit{S}_{\textrm{el}}}{\hbar } &\equiv& \int \; \frac{ dr d\tau}{2\pi u K} 
\left\{ \left[ \partial_{\tau} \phi(r,\tau) \right]^2 + u^2 \left[ \partial_{r} \phi(r,\tau) \right]^2 \right\}, 
\label{Eq:S_el}
\end{eqnarray}
with the imaginary time $\tau$ and the bosonic field 
$\phi(r,\tau)$.
 
Before continuing, we comment on  the Luttinger liquid parameter $K$. In existing experiments, its value is largely unknown, and only few attempts have been made to extract it experimentally.  In Ref.~\cite{Li:2015},  deduced values for InAs/GaSb 2DTIs are $K = 0.21$--0.22, indicating strong electron-electron interaction in that sample.~\footnote[5]{
On the other hand, we note that the value extracted for $K$ depends crucially on the theory to which the experiments are fitted~\cite{Vayrynen:2016,Hsu:2017}; see also the discussion in Sec.~\ref{Sec:hf-bs}.}
To be able to make quantitative predictions, we therefore use $K=0.2$ (see Table~\ref{Tab:parameters}) throughout this paper, unless stated otherwise. It represents the nuclear-spin-induced effects on the edge transport in the presence of strong electron--electron interactions.

The hyperfine interaction is given by  
\begin{equation}
\mathit{H}_{\textrm{hf}} = \frac{A_{0}}{\rho_{\textrm{nuc}}} \sum_{n} \left[ \Psi_{\textrm{el}}^{\dagger}({\bf x}_{n}) \frac{\boldsymbol{\sigma}}{2} \Psi_{\textrm{el}}({\bf x}_{n}) \right] \cdot {\bf I}_{n},
\label{Eq:H_hf}
\end{equation} 
which describes the coupling of an electron spin $\boldsymbol{\sigma} /2 \equiv (\sigma^{x}, \sigma^{y}, \sigma^{z})/2$ to nuclear spins ${\bf I}_{n}$ (with magnitude $I$) with the coupling constant $A_0$ at positions ${\bf x}_{n}$ with the nuclear index $n$. Here $\sigma^{\mu}$ denotes the $\mu$ component of the Pauli matrix vector in spin space. We take the nuclear density to be $\rho_{\textrm{nuc}} = 8/a_{0}^{3}$ with the lattice constant $a_{0}$~\cite{Meng:2014a,Kornich:2015}, and write the electron operator $\Psi_{\textrm{el}}$ as the product of the transverse (${\bf x_{\perp}}$) and the longitudinal ($r$) parts,
\begin{equation}
\Psi_{\textrm{el}}({\bf x})  = \psi_{\perp} ({\bf x_{\perp}}) \psi_{\shortparallel}(r).
\end{equation} 
In the above, the transverse part of the edge electron wave function $\psi_{\perp}$ is a complex scalar, while the longitudinal part is a two-component spinor, $\psi_{\shortparallel} = \left(L_{\uparrow}, R_{\downarrow}\right)^{\textrm{T}}$.  
For simplicity, we assume that the system is homonuclear and take the average values for the nuclear spin $I$ and the hyperfine coupling constant $A_{0}$. We also assume the electron wave function to be uniform in the transverse direction such that it is approximated as a constant $\left| \psi_{\perp} \right|^2 = 1 / (W_{\textrm{eff}}a)$, with $W_{\textrm{eff}}$ denoting the effective width of the wave function perpendicular to the 2DTI plane.\footnote{Assuming that the electrons are confined by a square potential well along the $z$ direction (perpendicular to the 2DTI plane), the inhomogeneous hyperfine coupling is then proportional to the electron density $\propto \sin^2 (\pi z/W_{\textrm{qw}})$ with the lithographic quantum well thickness $W_{\textrm{qw}}$. In order to incorporate such inhomogeneity, we define the effective width $W_{\textrm{eff}}$ by averaging the hyperfine coupling over $z$, such that $1/W_{\textrm{eff}} \equiv (1/W_{\textrm{qw}}) \int_{0}^{W_{\textrm{qw}}}dz \, \sin^2 (\pi z/W_{\textrm{qw}}) p(z)$, weighted by the probability distribution $p (z) = (2/W_{\textrm{qw}})\sin^2 (\pi z/W_{\textrm{qw}})$. As a result, we find that $W_{\textrm{eff}} = 4W_{\textrm{qw}}/3$, which is used to approximate the transverse electron wave function.} With these approximations, the hyperfine interaction can be written in a one-dimensional form,
\begin{equation}
\mathit{H}_{\textrm{hf}} \approx \frac{A_{0}}{N_\perp} \sum_{j} \boldsymbol{S}(r_{j}) \cdot \tilde{{\bf I}}_{j},
\label{Eq:H_hf_1d}
\end{equation}
with the number of nuclei per cross section $N_{\perp}=W_{\textrm{eff}}a a_{0} \rho_{\textrm{nuc}}$. Here $r_{j}$ denotes the common value of the edge coordinate of the nuclei belonging to the $j$-th cross section. In the above, we define the effective spin operator, 
\begin{equation}
\boldsymbol{S}(r) \equiv \frac{ N_{\perp} }{\rho_{\textrm{nuc}}} \left| \psi_{\perp} \right|^2 \sum_{\alpha,\beta} \psi_{\shortparallel,\alpha}^{\dagger} (r) \frac{\boldsymbol{\sigma}_{\alpha\beta}}{2} \psi_{\shortparallel,\beta} (r) ,
\end{equation}
which interacts with a classical spin $\tilde{{\bf I}}_{j} \equiv \sum_{n \in j} {\bf I}_{n}$ composed of $N_{\perp}$ nuclear spins within the $j$-th cross section. In Fig.~\ref{Fig:setup}, the composite spins $\tilde{{\bf I}}$ are plotted as the green arrows along the edges, in which the cross sections are drawn as yellow blocks, demonstrating the three-dimensionality of the nuclear subsystem.
A summary of the adopted parameters is given in Table~\ref{Tab:parameters}. We will examine various backscattering mechanisms arising from Eq.~(\ref{Eq:H_hf_1d}), and their contributions to the edge resistance.

As a remark, the dipolar interaction between the nuclear spins is much weaker than the hyperfine interaction~\cite{Paget:1977}, and is not explicitly included in the above. In the disordered phase, however, it leads to the dissipation of the accumulated nuclear spin polarization during the backscattering process due to the accompanied electron-nuclear flip-flops~\cite{Lunde:2012,Kornich:2015,Russo:2017}. In order to incorporate the effect of the dipolar interaction, we assume that the nuclear spin polarization is destroyed by such a dissipation channel, and adopt unpolarized nuclear spin orientation for our analysis in the disordered phase [see Eq.~(\ref{Eq:random})]. On the other hand, the RKKY interaction between the nuclear spins dominates the dipolar interaction, so the latter is neglected in our analysis of the nuclear spin ordering. 
In addition, the Kondo temperature associated with a single nuclear spin is $T_{\textrm{Kondo}}=\Delta e^{-\epsilon_{F}/A_{0}}$~\cite{Simon:2007}, so the Kondo physics should not be relevant, as $\epsilon_{F} \gg A_{0}$ under typical experimental conditions (see also Refs.~\cite{Maciejko:2012,Schecter:2015,Yevtushenko:2017} for discussions on the limitations of the RKKY description).

\begin{table*}[t]
\caption{Physical parameters and estimated quantities for InAs/GaSb quantum wells, HgTe/(Hg,Cd)Te quantum wells, and GaAs nanowires. 
We also include the nonhelical edge channels in the trivial regime of the InAs/GaSb quantum wells.}
\begin{tabular}[c]{  l  c  c  c  c }
\hline \hline
Physical parameter & InAs/GaSb~\footnote{From Refs.~\cite{Gueron:1964,Paget:1977,Schliemann:2003,Wu:2006,Braun:2006,Maciejko:2009,Knez:2011,Pribiag:2015,Schrade:2015,Li:2015}.} & HgTe/(Hg,Cd)Te~\footnote{From Refs.~\cite{Konig:2007,Roth:2009,Hou:2009,Maciejko:2009,Strom:2009,Teo:2009,Egger:2010,Lunde:2013}.}  & InAs/GaSb (trivial)~\footnote{For the trivial regime of InAs/GaSb, we take the same parameters as in the footnote~a except that in this case there are two Luttinger liquid parameters $K_{c}$ and $K_{s}$ for the two conducting channels.} & GaAs~\footnote{From Refs.~\cite{Paget:1977,Pfeiffer:1997,Yacoby:1997,Auslaender:2002,Auslaender:2005,Schliemann:2003,Braun:2006,Steinberg:2008,Braunecker:2009a,Braunecker:2009b}.}  \\
\hline
Hyperfine coupling constant, $A_{0}$ & 50~$\mu$eV & 3~$\mu$eV~\footnote{Here we take the arithmetic average of the hyperfine coupling over all the nuclei, weighted by their natural abundance. The average value of HgTe/(Hg,Cd)Te is small because only $\sim$19\% of the naturally abundant nuclei in this material possess nonzero spins.} & 50~$\mu$eV  & 90~$\mu$eV  \\
Nuclear spin, $I$ & 3~\footnote{The given value is an approximate average over the stable isotopes, defined by $I(I+1) \equiv \sum_{m} \rho_{m}^{\textrm{iso}} I_{m}(I_{m}+1)$ with the natural abundance $\rho^{\textrm{iso}}$ and the index $m$ labeling the isotopes. We used $I(^{113}\textrm{In})=I(^{115}\textrm{In})=9/2$, $I(^{75}\textrm{As})=3/2$, $I(^{69}\textrm{Ga})=I(^{71}\textrm{Ga})=3/2$, $I(^{121}\textrm{Sb})=5/2$ (with $\rho^{\textrm{iso}} \sim 57\%$), and $I(^{123}\textrm{Sb})=7/2$ (with $\rho^{\textrm{iso}} \sim 43\%$).} & 0.3 ~\footnote{Similarly as for the footnote~f. Here we used $I(^{123}\textrm{Te})=I(^{125}\textrm{Te})=I(^{111}\textrm{Cd})=I(^{113}\textrm{Cd})=1/2$, $I(^{199}\textrm{Hg})=1/2$ (with $\rho^{\textrm{iso}} \sim 17\%$), and $I(^{201}\textrm{Hg})=3/2$ (with $\rho^{\textrm{iso}} \sim 13\%$).} & 3  & 3/2 \\
Fermi velocity, $v_{F}$ & $4.6 \times 10^4~$m/s & $5.1 \times 10^5~$m/s & $4.6 \times 10^4~$m/s  & $2.0 \times 10^5~$m/s  \\ 
Fermi wave number, $k_{F}$ & $7.9 \times 10^{7}~$m$^{-1}$ & $7.3 \times 10^{7}~$m$^{-1}$ & $7.9 \times 10^{7}~$m$^{-1}$  & $1.2 \times 10^{8}~$m$^{-1}$ \\
Lattice constant, $a_{0}$ & 6.1~\AA & 6.5~\AA & 6.1~\AA & 5.7~\AA \\
Transverse decay length, $a$ & 9~nm & 14~nm & 9~nm & -- \\
Quantum well width, $W_{\textrm{qw}}$ & 15~nm & 9~nm & 15~nm  & -- \\
Cross section area & 9 $\times$ 15~nm$^2$ & 14 $\times$ 9~nm$^2$ & 9 $\times$ 15~nm$^2$  & 10 $\times$ 10~nm$^2$ \\
Number of nuclei per cross section, $N_{\perp}$ & 3900 & 3200 & 3900  & 2500 \\
Bulk gap, $\Delta=\hbar v_{F} / a$  & 3.4~meV & 24~meV & 3.4~meV  & -- \\
Bandwidth, $\Delta_{a}=\hbar v_{F} / a_{0}$  & -- & -- & --  & 0.23~eV \\
Luttinger liquid parameter(s)  & $K=0.2$ & $K=0.2$ & $K_{c}=0.2$, $K_{s}=1$  & $K_{c}=0.2$, $K_{s}=1$ \\
Mean free path, $\lambda_{\textrm{mfp}}$ & 0.1--1~$\mu$m & 0.1--1~$\mu$m & 0.1--1~$\mu$m  & 0.1--1~$\mu$m \\
\hline \hline
Estimated quantity & & & & \\
\hline
Backscattering on disordered nuclear spins & & & & \\
\hspace{5pt} Localization length, $\xi_{\textrm{hf}}$ & 17~$\mu$m & 3.7~mm & 42~$\mu$m  & 0.17~mm \\ 
\hspace{5pt} Localization temperature, $T_{\textrm{hf}}$ & 100~mK  & 5.3~mK & 19~mK  & 20~mK \\
\hspace{5pt} Electronic gap, $\Delta_{\textrm{hf}}$ & 1.2~$\mu$eV & 64~neV & 0.79~$\mu$eV  & 0.86~$\mu$eV \\
Nuclear spin ordering & & & & \\
\hspace{5pt} Transition temperature, $T_{0}$ & 42~mK & 1.4~mK & 35~mK  & 29~mK \\
\hspace{5pt} Electronic gap at $T=0$, $\Delta_{\textrm{m}}(T=0)$ & 0.36~meV & 50~$\mu$eV & 1.2~meV  & 2.1~meV \\
Nuclear-spin-order-assisted backscattering on impurities & & & & \\
\hspace{5pt} Localization length at $T=0$, $\xi_{\textrm{hx}}(T=0)$ &  7.9--19~$\mu$m & 3.2--7.7~mm & 0.93-2.5~$\mu$m  & 4.7--13~$\mu$m \\ 
\hspace{5pt} Characteristic temperature, $T_{\textrm{hx}}$ & 92--220~mK & 2.5--6.1~mK & 0.32--0.85~K  & 0.27--0.72~K \\
\hspace{5pt} Electronic gap at $T=0$, $\Delta_{\textrm{hx}}(T=0)$ & 1.1--2.7~$\mu$eV & 31--74~neV & 10--27~$\mu$eV  & 8.7--23~$\mu$eV \\
\hline \hline
\end{tabular}
\label{Tab:parameters}
\end{table*}

\section{Elastic backscattering on randomly oriented nuclear spins~\label{Sec:hf-bs}}

We first consider the disordered phase, where the nuclear spins are randomly oriented, and cause elastic electron backscattering via the hyperfine interaction Eq.~(\ref{Eq:H_hf_1d}). 
Since the forward scattering [the $S^{z}$ term in Eq.~(\ref{Eq:H_hf_1d})] has no influence on the transport properties~\cite{Giamarchi:1988,Giamarchi:2003}, it can be dropped. In the continuum limit, the remaining components of the electron spin operator can be written in terms of the right and left movers, 
\begin{subequations}
\begin{eqnarray}
S^{x}(r) &=& \frac{1}{2} \left[ L_{\uparrow}^{\dagger} (r) R_{\downarrow} (r) + R_{\downarrow}^{\dagger}(r) L_{\uparrow}(r) \right], \\ 
S^{y}(r) &=& \frac{-i}{2} \left[ L_{\uparrow}^{\dagger} (r) R_{\downarrow} (r) - R_{\downarrow}^{\dagger}(r) L_{\uparrow}(r) \right], 
\end{eqnarray}
\end{subequations}
and then bosonized using Eq.~(\ref{Eq:bosonization}). This gives rise to the backscattering Hamiltonian,
\begin{eqnarray}
\mathit{H}_{\textrm{hf,b}} &=& \int \frac{dr}{2\pi a} V_{\textrm{hf},2k_{F}} (r) e^{2i \phi (r) } + \textrm{H.c.}, 
\label{Eq:H_hf_b}
\end{eqnarray}
where we keep only the slowly varying terms, and define the $2k_{F}$ component of the random potential caused by the nuclear spins as
\begin{eqnarray}
V_{\textrm{hf},2k_{F}} (r) &\equiv& \frac{A_{0}}{2 N_{\perp}} \left[ \tilde{I}^{x} (r) + i \tilde{I}^{y} (r) \right] e^{-2i k_{F} r}.
\end{eqnarray}
Assuming that the nuclear spins are independent and unpolarized, we have 
\begin{eqnarray}
\left\langle V_{\textrm{hf},2k_{F}}^{\dagger} (r) V_{\textrm{hf},2k_{F}} (r') \right\rangle_{\textrm{rs}} &=& M_{\textrm{hf}} \delta(r-r'), 
\label{Eq:random}
\end{eqnarray}
with the strength $M_{\textrm{hf}} \equiv a A_{0}^2 I(I+1)/(6 N_{\perp})$. Here $\left\langle \cdots \right\rangle_{\textrm{rs}}$ denotes the expectation value with respect to the random nuclear spin state. Using the replica method~\cite{Giamarchi:2003} and the average from Eq.~(\ref{Eq:random}), we obtain the contribution to the imaginary-time action from Eq.~(\ref{Eq:H_hf_b}), 
\begin{eqnarray}
\frac{\delta \mathit{S}_{\textrm{hf}}}{\hbar} 
&=& - \frac{D_{\textrm{hf}} u^2}{8\pi a^3 }  \int_{u|\tau-\tau'|>a}  dr d\tau d\tau' \; \nonumber \\
&& \hspace{0.8in} \times \cos \left[  2 \phi (r,\tau)- 2\phi (r,\tau') \right],
\label{Eq:S_hf-bs}
\end{eqnarray}
with the dimensionless coupling constant $D_{\textrm{hf}} \equiv 2a M_{\textrm{hf}} /(\pi \hbar^{2} u^{2})$. With the effective action composed of Eqs.~(\ref{Eq:S_el}) and (\ref{Eq:S_hf-bs}), we perform the RG analysis to investigate how the disordered nuclear spins affect the transport properties of the edge states.

Following the procedure in Appendix~\ref{Appendix:RG}, we build the RG flow equations by first evaluating the correlation function,
\begin{equation}
\left< e^{i [\phi({\bf r_{1}}) - \phi({\bf r_{2}})] } \right>_{\mathit{S}_{\textrm{el}}+\delta \mathit{S}_{\textrm{hf}}},
\end{equation}
with respect to Eqs.~(\ref{Eq:S_el}) and (\ref{Eq:S_hf-bs}). Then, upon changing the cutoff $a\rightarrow a(l) = ae^{l}$ with the dimensionless scale $l$, we find the RG flow equations,
\begin{subequations}
\label{Eq:RG_hf}
\begin{eqnarray}
\frac{d D_{\textrm{hf}} (l) }{d l } &=& \left[ 3-2K(l) \right] D_{\textrm{hf}} (l), \\
\frac{d K (l) }{d l } &=& - \frac{K^2 (l) }{2} D_{\textrm{hf}}(l), \\
\frac{d u (l) }{d l } &=& -\frac{u(l) K (l)}{2} D_{\textrm{hf}}(l),
\end{eqnarray}
\end{subequations}
from which we see that the backscattering on disordered nuclear spins Eq.~(\ref{Eq:S_hf-bs}) is RG relevant for $K(l)<3/2$. 
Assuming that the change in $K$ can be neglected, we integrate the RG flow of the effective coupling to get $D_{\textrm{hf}} (l)= D_{\textrm{hf}} (l=0) e^{(3-2K)l}$, which allows us to find the localization length,
\begin{equation}
\xi_{\textrm{hf}} = a D_{\textrm{hf}}^{-1/(3-2K)}, 
\end{equation}
and the corresponding localization temperature $T_{\textrm{hf}}\equiv \hbar u / (k_{B} \xi_{\textrm{hf}})$. For an edge longer than $\xi_{\textrm{hf}}$, the conductance gets exponentially suppressed below $T_{\textrm{hf}}$. For the parameters of InAs/GaSb 2DTIs (see Table~\ref{Tab:parameters}), the estimated values of $\xi_{\textrm{hf}}$ and $T_{\textrm{hf}}$ suggest that the localization-delocalization transition is within an experimentally accessible regime. In contrast, the spinful nuclei in HgTe/(Hg,Cd)Te are naturally less  abundant, possess smaller spins, and have weaker hyperfine coupling~\cite{Lunde:2013}, leading to a much bigger localization length and much lower localization temperature. 
 
\begin{figure}[t]
\centering
\includegraphics[width=\linewidth]{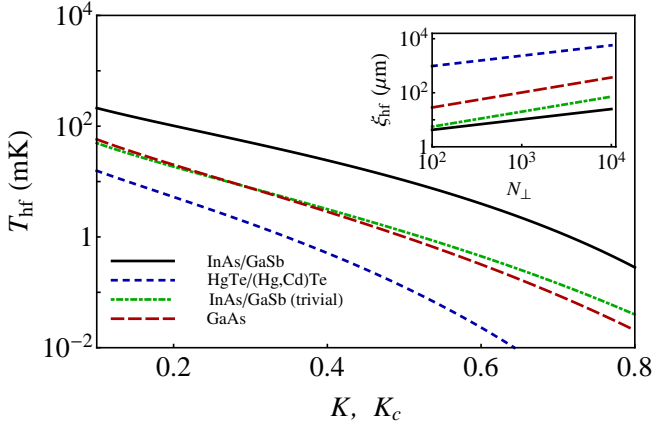}
\caption{Dependence of the localization temperature $T_{\textrm{hf}}$ on strength of electron-electron interactions for various materials. For the horizontal axis, we take the Luttinger liquid parameter $K$ for 2DTI helical edge states, and $K_{c}$ (while fixing $K_{s}=1$) for nonhelical channels in GaAs wires and in the trivial regime of InAs/GaSb. The other parameters are listed in Table~\ref{Tab:parameters}. Inset: the localization length $\xi_{\textrm{hf}}$ as a function of the number of nuclei per cross section $N_{\perp}$.}
\label{Fig:T_hf_K}
\end{figure}

Since the Luttinger liquid parameter $K$ and the number of nuclei per cross section $N_{\perp}$ vary with materials and setups, we also investigate the dependence of the localization temperature and length on these parameters. In Fig.~\ref{Fig:T_hf_K} we plot the localization temperature of InAs/GaSb and HgTe/(Hg,Cd)Te 2DTIs, in addition to nonhelical states in the trivial regime of InAs/GaSb and a spin-degenerate quasi-one-dimensional GaAs wire, as a function of the interaction parameter. The localization temperatures of these materials drastically increase with stronger interactions, a feature that can be tested through the sample preparation, e.g. by varying the quantum well width, or the distance between a screening metallic gate and the quasi-one-dimensional channel. 
In addition to the difference between the InAs/GaSb and the HgTe/(Hg,Cd)Te 2DTIs due to material parameters, there is a pronounced difference in the localization temperatures between a helical edge of InAs/GaSb, and nonhelical channels in the trivial regime of InAs/GaSb and a GaAs conductor, in spite of their comparable nuclear spins and hyperfine couplings. This difference arises from the fact that in a spin-degenerate wire the effective Luttinger liquid parameter is $K_{\textrm{wire}} = (K_{c} + 1/K_{s})/2$, an average of values for the charge ($K_{c}$) and spin ($K_{s} \approx 1$) channels, and thus bounded by $1/2$, whereas such averaging is absent in a helical edge. 
Consequently, in 2DTIs the interaction leads to a stronger effect on the nuclear-spin-induced localization, which may reveal the helical nature of the edge states in the band-inverted regime of InAs/GaSb. The dependence of $\xi_{\textrm{hf}}$ on the interaction can be inferred from Fig.~\ref{Fig:T_hf_K}, using the fact that $\xi_{\textrm{hf}}$ is inversely proportional to $T_{\textrm{hf}}$, and therefore not displayed here.
In the inset of Fig.~\ref{Fig:T_hf_K}, we plot the dependence of $\xi_{\textrm{hf}}$ on $N_{\perp}$, which depends on the quantum well width and the transverse decay length, and can also vary for different samples. Since the dependence of $\xi_{\textrm{hf}}$ (and therefore $T_{\textrm{hf}}$) on $N_{\perp}$ is a fractional power law, its value does not change much even if $N_{\perp}$ varies by an order of magnitude. From now on we shall adopt the parameters of InAs/GaSb 2DTIs, in which we expect the most significant effects from the nuclear spins.

Before continuing, we note that other possible backscattering mechanisms may cause edge resistance, and therefore contribute to the total resistance $R_{\textrm{total}}$, in addition to the contact resistance from the leads and the nuclear-spin-induced resistance $R$. 
However, since here we want to find out whether the nuclear-spin-induced resistance is observable, i.e. whether $R$ is comparable to the resistance quantum $R_{0} \equiv h/e^2$, throughout the paper we discuss and plot $R/R_{0}$, instead of $R_{\textrm{total}}$.

\begin{figure}[t]
\centering
\includegraphics[width=\linewidth]{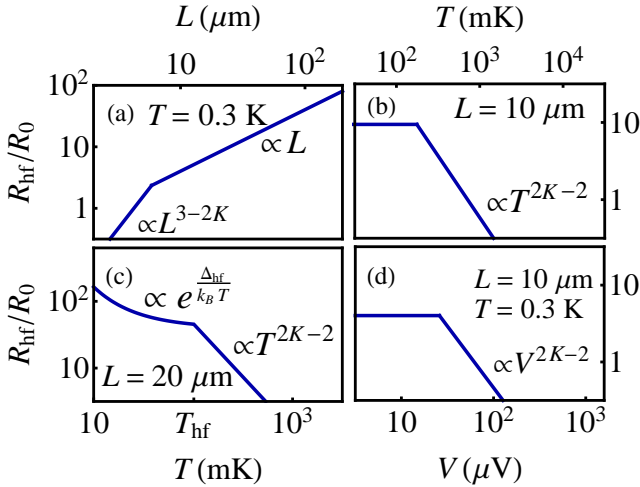}
\caption{
Dependence of the resistance $R_{\textrm{hf}}$ on different system parameters in the disordered phase for the parameters of InAs/GaSb. (a) $R_{\textrm{hf}}$ versus length ($L$) at the temperature $T=0.3~$K. (b) $R_{\textrm{hf}}$ versus $T$ for $L=10~\mu$m. (c) $R_{\textrm{hf}}$ versus $T$ for $L=20~\mu$m. (d) Differential resistance as a function of the bias voltage ($V$) for $L=10~\mu\textrm{m}$ at $T=0.3~$K. 
In all the panels we take the Luttinger liquid parameter $K=0.2$. 
The other parameters are listed in Table~\ref{Tab:parameters}.
}
\label{Fig:R_hf}
\end{figure}

We now investigate the resistance $R_{\textrm{hf}}$ caused by disordered nuclear spins, Eq.(\ref{Eq:S_hf-bs}). Using the effective coupling $D_{\textrm{hf}} (l)$, we compute the edge conductivity as in Refs.~\cite{Giamarchi:1988,Giamarchi:2003}, and therefore the edge resistance,
\begin{eqnarray}
R_{\textrm{hf}} & \propto & R_{0} \frac{M_{\textrm{hf}} L }{\hbar^2 v_{F}^2} e^{(2-2K)l^{*}}.
\end{eqnarray} 
Here the dimensionless scale $l^{*}$ arises from the cutoff $a(l^{*}) = a e^{l^{*}}$, at which the RG flow stops, and thus depends on the experimental conditions. We identify four possible physical cutoffs, being the edge length $L$, the thermal length $\lambda_{T}\equiv \hbar u / (k_{B} T)$, the bias length $\lambda_{V}\equiv \hbar u / (e V)$, and the localization length $\xi_{\textrm{hf}}$, corresponding to the short-edge, the high-temperature, the high-bias, and the strong-coupling regimes, respectively. 
First, if the edge length is the shortest among all these scales, $L< \lambda_{T},~\lambda_{V},~\xi_{\textrm{hf}}$, we obtain
\begin{eqnarray}
R_{\textrm{hf}} (L) &\propto& R_{0} \frac{\pi D_{\textrm{hf}}}{2K^2} \left( \frac{L}{a} \right)^{3-2K}.
\label{Eq:R_hf_L}
\end{eqnarray}
Second, if the temperature is so high that $\lambda_{T}<L,~\lambda_{V},~\xi_{\textrm{hf}}$, we get
\begin{eqnarray} 
R_{\textrm{hf}} (T) & \propto & R_{0} \frac{\pi D_{\textrm{hf}} L }{2K^2 a} \left( \frac{K k_{B}T}{\Delta} \right)^{2K-2}.
\label{Eq:R_hf_T}
\end{eqnarray} 
Third, at high bias such that $\lambda_{V} < L,~\lambda_{T},~\xi_{\textrm{hf}}$, the differential resistance depends on the bias voltage as 
\begin{eqnarray}
\left( \frac{dV}{dI} \right)_{\textrm{hf}} & \propto & R_{0} \frac{\pi D_{\textrm{hf}} L }{2K^2 a} \left( \frac{K eV}{\Delta} \right)^{2K-2}.
\label{Eq:R_hf_V}
\end{eqnarray} 
Finally, if $\xi_{\textrm{hf}} < L,~\lambda_{T},~\lambda_{V}$, the RG flow reaches the strong-coupling regime, so the edge states are gapped, displaying a thermally activated resistance~\cite{Giamarchi:2003},
\begin{eqnarray}
R_{\textrm{hf}} (T) &\propto& R_{0} \frac{\pi D_{\textrm{hf}}L }{2K^2 a}  e^{\Delta_{\textrm{hf}} /(k_{B}T)},
\label{Eq:R_hf_gapped}
\end{eqnarray}
with the gap $\Delta_{\textrm{hf}} = \Delta \left( 2K D_{\textrm{hf}} \right)^{1/(3-2K)}$. The formulas Eqs.~(\ref{Eq:R_hf_L})--(\ref{Eq:R_hf_gapped}) were given in Ref.~\cite{Hsu:2017} (with slightly different notations), and are repeated here for reference. 
Further, here we additionally check that the edge resistance due to the disordered nuclear spins in the trivial regime of InAs/GaSb is much smaller than the one in the topological phase (by two orders of magnitudes, not shown), consistent with our conclusion from Fig.~\ref{Fig:T_hf_K}.
We also note that in the localized regime, the temperature dependence of the resistance may be affected by the tunneling between the instanton/kink states. For nonhelical systems, the conductivity/conductance due to such tunneling events has been investigated~\cite{Nattermann:2003,Aseev:2017}. Since it is beyond the scope of this paper, here we only note that the variable-range-hopping behavior $\sigma (T) \propto $~Exp[$-\sqrt{C/(k_{B}T)}$] with a constant $C$ due to the tunneling between the kink states in the localized regime~\cite{Nattermann:2003} may be relevant to the observation in Ref.~\cite{Nichele:2016}. 

Figure \ref{Fig:R_hf} summarizes the dependence of the resistance on the most relevant and accessible parameters, as given, depending on the regime, by Eqs.~(\ref{Eq:R_hf_L})--(\ref{Eq:R_hf_gapped}). 
In panel (a) of Fig.~\ref{Fig:R_hf}, we show how the resistance scales with the edge length. The kink in the curve signifies the transition from the $L^{3-2K}$ power law for a short edge [Eq.~\eqref{Eq:R_hf_L}] to the linear $L$ dependence for a long edge [Eq.~\eqref{Eq:R_hf_T}]. 
The temperature dependence is shown in panels (b) and (c). The resistance initially increases as a $T^{2K-2}$ upon decreasing the temperature $T$. After that the resistance for a short edge $L<\xi_{\textrm{hf}}$ saturates [panel (b)] because the RG flow stops at the edge length. In contrast, for a long edge $L>\xi_{\textrm{hf}}$ it evolves into an exponential [panel (c)] due to the electronic gap in the spectrum. As a result, for an edge of the length longer than $\xi_{\textrm{hf}}$, the localization of the edge states caused by disordered nuclear spins is observable below the localization temperature $T_{\textrm{hf}}$. Finally, the differential resistance in the presence of a finite bias voltage $V$ is plotted in the panel (d) for an edge shorter than $\xi_{\textrm{hf}}$. Starting from the high-bias regime, the differential resistance initially increases with a decreasing voltage as a power law [Eq.~\eqref{Eq:R_hf_V}], and then saturates due to the cutoff given by the shorter of $L$ and $\lambda_{T}$, Eqs.~\eqref{Eq:R_hf_L} and \eqref{Eq:R_hf_T}, respectively. 

We conclude here that the power law dependencies of the resistance are symptomatic for our theory.
Recently, such fractional power laws were reported~\cite{Li:2015} in short InAs/GaSb 2DTI samples as a function of the temperature and the bias voltage.
Similar measurements with longer samples can be therefore used to examine and distinguish various theories, including ours, proposed for the origin of the edge resistance. 
For such comparison, an independently extracted value of the parameter $K$ for the edge states~\cite{Ilan:2012,Muller:2017} would be highly desirable.

\section{Spiral nuclear spin order~\label{Sec:ordering}}

In addition to the backscattering effects, the interplay between the nuclear spins and the strong electron-electron interaction leads to the formation of a spiral nuclear spin order, which will be discussed in this section.

\subsection{RKKY interaction and spiral nuclear spin order~\label{SubSec:RKKY}}

We now discuss the nuclear spin order stabilized by the edge electron-mediated RKKY interaction. Since the energy scale of the hyperfine coupling is much smaller than the Fermi energy, we can integrate out the electron degrees of freedom in the hyperfine interaction defined in Eq.~(\ref{Eq:H_hf_1d}) to obtain the RKKY interaction, a pairwise coupling between the static nuclear spins~\cite{Simon:2007,Simon:2008,Braunecker:2009a,Braunecker:2009b,Klinovaja:2013a,Klinovaja:2013b,Meng:2014a,Hsu:2015,Yang:2016},   
\begin{eqnarray}
\mathit{H}_{\textrm{RKKY}} &=& \frac{1}{N_{\perp}^{2}} \sum_{i,j,\mu} J^{\mu}_{ij} \tilde{I}_{i}^{\mu} \tilde{I}_{j}^{\mu},
\label{Eq:H_RKKY} 
\end{eqnarray}
with $\mu=x,y,z$ in the spin space. Here the RKKY coupling $J^{\mu}_{ij}$ is proportional to the electronic spin susceptibility, and can be calculated along the line of Ref.~\cite{Giamarchi:2003}. Since the $z$ component of the electron spin operator is marginally relevant in the helical Tomonaga-Luttinger liquid, the $z$ component of the RKKY coupling $J^{z}_{ij}$ decays as $1/(r_{i}-r_{j})^{2}$~\cite{Giamarchi:2003}, negligible compared to the $x$ and $y$ components. This is a consequence of the broken SU(2) spin rotational symmetry of the edge states, and it leads to an anisotropic RKKY coupling $|J^{x}_{q}|=|J^{y}_{q}| \gg |J^{z}_{q}|$ in momentum space, where the $x$ and $y$ components of the RKKY coupling are given by
\begin{align}
 J^{x}_{q} =& J^{y}_{q} \nonumber \\
 =& - \frac{  \sin (\pi K) }{8\pi^2} \frac{K A_{0}^2}{ \Delta }
\left(\frac{\lambda_{T}}{2\pi a} \right)^{2-2K} \label{Eq:JRKKY}\\
&\times \sum_{\kappa=\pm}
\left| \frac{ \Gamma\left( 1-K\right) \Gamma\left[  K / 2-i \lambda_{T} \left( q - 2 \kappa k_{F} \right)/ (4\pi) \right]}
{\Gamma\left[ (2-K)/2 - i \lambda_{T} \left( q - 2 \kappa k_{F} \right)/ (4\pi) \right]}
\right|^{2}, 
\nonumber
\end{align}
with the Gamma function $\Gamma(x)$. In addition to the anisotropy, the helicity of the electrons also leads to a stronger RKKY coupling, compared to the nonhelical case, because of the difference in the effective Luttinger liquid parameters ($K_\textrm{wire}$ versus $K$), as explained in Sec.~\ref{Sec:hf-bs}.
 
The RKKY coupling given by Eq.~(\ref{Eq:JRKKY}) develops a dip at $q= \pm 2k_{F}$, and therefore gives rise to an instability toward a nuclear spin order in a finite-size system.
Even though a similar RKKY-induced nuclear spin order also arises in nonhelical, spin-degenerate systems such as GaAs quantum wires and $^{13}$C nanotubes~\cite{Braunecker:2009a,Braunecker:2009b,Klinovaja:2013b,Meng:2013,Scheller:2014,Meng:2014a,Stano:2014,Hsu:2015}, we note four important differences regarding to the nuclear orders between a helical edge and a nonhelical wire. 

First, the ordered nuclear spins align ferromagnetically within each cross section. Along the edge ($x$ axis), they rotate in the $xy$ plane with a spatial period $\pi/k_F$. For illustration, the nuclear spin order is displayed in Fig.~\ref{Fig:setup}.
The plane within which the nuclear spins rotate is fixed by the 2DTI plane, due to the broken SU(2) symmetry in the edge states. This is different to the nuclear spin helix formed in a spin-degenerate system~\cite{Braunecker:2009a,Braunecker:2009b,Klinovaja:2013b,Meng:2014a,Hsu:2015}, where the nuclear spins rotate in a plane which can have arbitrary orientation. 

Second, the remaining U(1) symmetry in a helical edge, corresponding to the rotation of nuclear spins around the spin quantization ($z$) axis, leads to one Goldstone mode in the magnon spectrum in an infinitely long system (cf. below). This is in contrast to nonhelical systems, where multiple Goldstone modes associated with the SU(2) symmetry emerge in the magnon spectrum~\cite{Braunecker:2009a,Braunecker:2009b,Klinovaja:2013b,Meng:2014a,Hsu:2015}.
 
Third, the tendency toward the nuclear spin order is typically higher for a 2DTI edge, as a result of the stronger RKKY coupling. This is essentially due to, again, the difference in the effective Luttinger liquid parameters, and it leads to a higher transition temperature $T_{0}$ [see Eq.~(\ref{Eq:T0}) and Table~\ref{Tab:parameters}]. In other words, the helical nature of the edge states promotes the formation of the nuclear spin order.

Finally, the nuclear spin ground state is related to the helicity of the electronic subsystem in a helical edge, unlike in a spin-degenerate wire. The expectation value of the nuclear spins in the ordered phase is given by
\begin{equation}
\left\langle {\bf \tilde{I}}(r) \right\rangle_{\pm} = N_{\perp} I m_{2k_{F}}  \, \left[ \cos (2k_{F} r) \hat{x} \pm \sin (2k_{F} r) \hat{y} \right],
\label{Eq:order}
\end{equation}
with the sign $\pm$ labeling the anticlockwise/clockwise rotation of the nuclear spins, and $m_{2k_{F}}$ denoting the order parameter such that $m_{2k_{F}}(T=0)=1$ for a complete order. In a nonhelical wire, both orders with the $\pm$ signs can be the ground state. As the temperature is lowered below $T_{0}$, the nuclear spins form an order with one of them (within a magnetic domain).
The ordered nuclear spins then generate a macroscopic Overhauser field, which acts back on the electron spins. Depending on the sign, either the conduction modes $R_{\downarrow}$ and $L_{\uparrow}$, or the other subbands are gapped out~\cite{Braunecker:2009a,Braunecker:2009b}. 
The nuclear spin helix in a nonhelical wire thus leads to a partial gap at the Fermi surface, halving the conductance~\cite{Braunecker:2009a,Braunecker:2009b,Scheller:2014,Aseev:2017}.

\begin{figure}[t]
\centering
\includegraphics[width=0.48\linewidth]{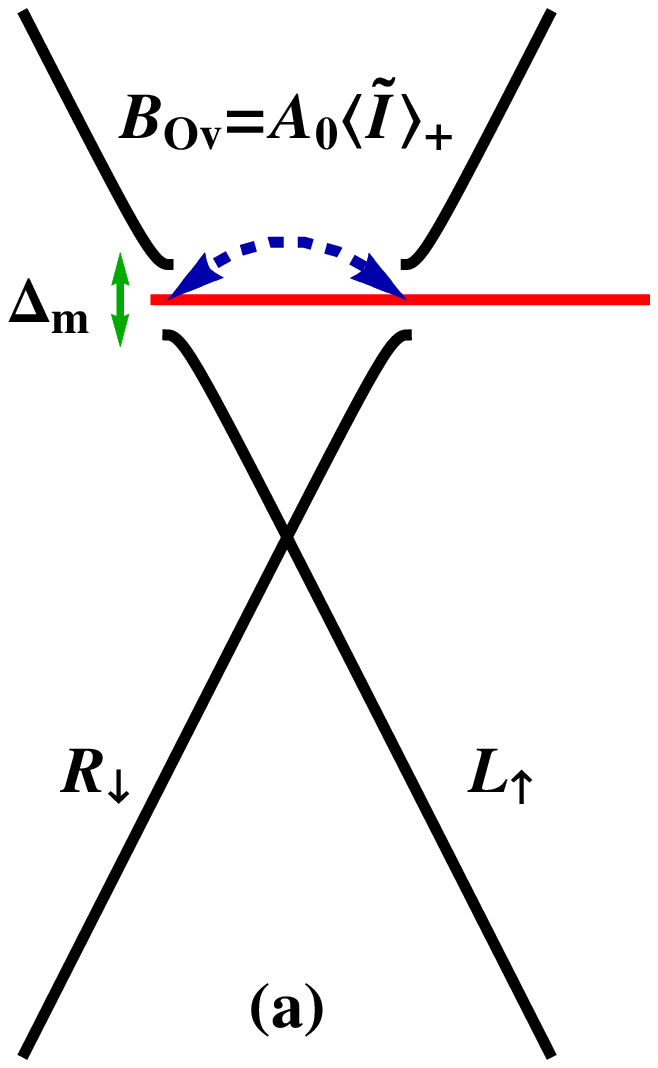}
\hspace{0.04\linewidth}
\includegraphics[width=0.46\linewidth]{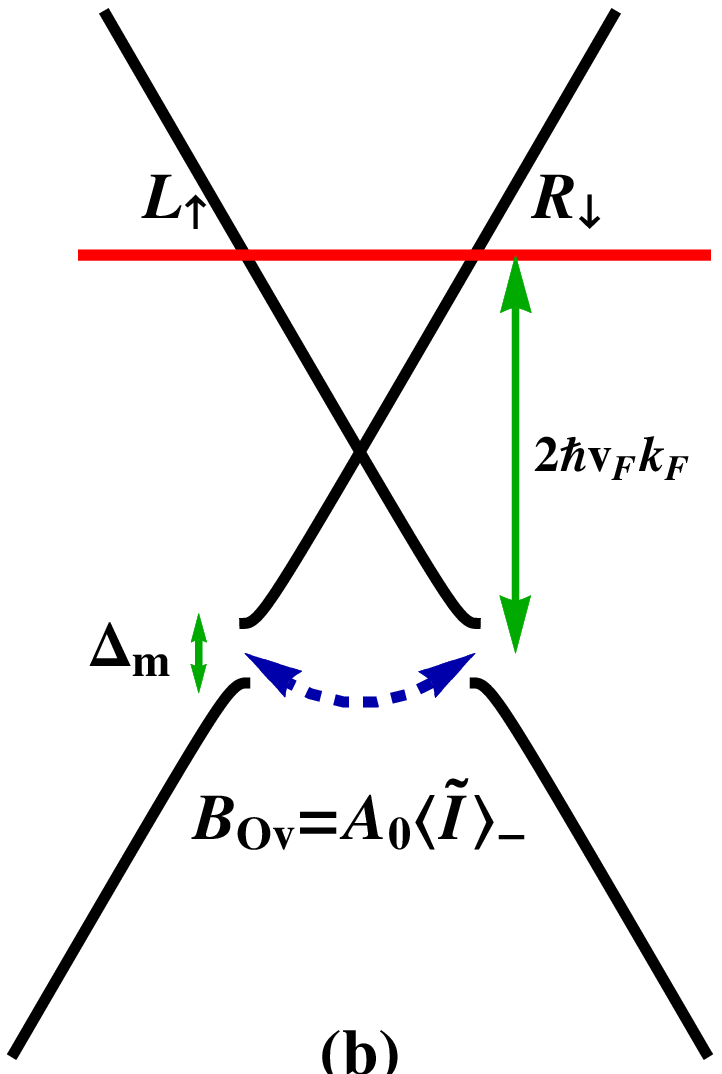}
\caption{The effect of nuclear helix $\langle {\bf \tilde{I}} \rangle_{\pm}$ on the electron spectrum. (a) For the spiral order $\langle {\bf \tilde{I}} \rangle_{+}$, a gap of the size $\Delta_{\textrm{m}}$ is opened at the Fermi surface (indicated by red line) by the corresponding Overhauser field. (b) For $\langle {\bf \tilde{I}} \rangle_{-}$, the gap is induced below the Fermi surface.}
\label{Fig:helicity}
\end{figure}

In a helical edge, however, the position of the electronic gap opened in the edge state spectrum depends on the $\pm$ sign in Eq.~(\ref{Eq:order}). Provided that the edge states consist of $R_{\downarrow}$ and $L_{\uparrow}$ electrons and the Fermi level is placed above the Dirac point at $q=0$, the order with the positive sign, $\langle {\bf \tilde{I}} \rangle_{+}$, would mix $R_{\downarrow}(q+2k_{F})$ and $L_{\uparrow}(q)$, and therefore gap out the electrons at the Fermi surface [panel (a) of Fig.~\ref{Fig:helicity}], reducing the RKKY coupling. On the other hand, $\langle {\bf \tilde{I}} \rangle_{-}$ would mix $R_{\downarrow}(q)$ and $L_{\uparrow}(q+2k_{F})$. In this case, a gap $\Delta_{\textrm{m}}$ is induced below the Fermi surface, as shown in panel (b) of Fig.~\ref{Fig:helicity}.
By establishing the nuclear spin order, the entire system of the nuclei and the electrons may acquire the magnetic energy gain, in addition to the Peierls energy (from opening an electronic gap at the Fermi surface) and the Knight energy (from the electron spin polarization)~\cite{Meng:2014a}.  Hence, the orders with the opposite signs lead to distinct energy gains due to the different gap positions. 
We examine the two scenarios [positive versus minus signs in Eq.~(\ref{Eq:order})], and find that the total energy gain of $\langle {\bf \tilde{I}} \rangle_{-}$ is higher, due to the stronger RKKY coupling and therefore the magnetic energy gain is larger (typically, the magnetic energy dominates the Peierls and Knight energies).
As a consequence, it is energetically favorable for the nuclear spins to order {\it without} opening a gap at the Fermi surface. 
To distinguish from the order in a nonhelical wire, which respects distinct symmetries, the order in a 2DTI edge predicted in this work is thus dubbed a {\it spiral} nuclear spin order.

\begin{figure}[t]
\centering
\includegraphics[width=\linewidth]{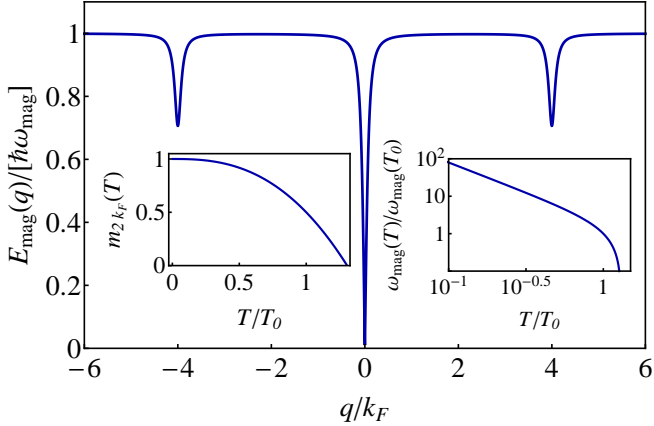}
\caption{Magnon energy $E_{\textrm{mag}}$ as a function of the momentum $q$ in an infinitely long edge. We take the order parameter $m_{2k_{F}}=1$ and an exaggerated temperature $T=10~$K to make the dip at $q=0$ visible. For a realistic temperature, the dips are very narrow, and the magnon energy is approximately momentum independent $E_{\textrm{mag}}(q) \approx \hbar \omega_{\textrm{mag}}$. The $T$ dependence of $m_{2k_{F}}$ ($\hbar\omega_{\textrm{mag}}$) given by Eq.~(\ref{Eq:Bloch}) [by Eq.~(\ref{Eq:omega_mag_T})] is shown in the left (right) inset.}
\label{Fig:magnon}
\end{figure}

\begin{figure}[t]
\centering
\includegraphics[width=\linewidth]{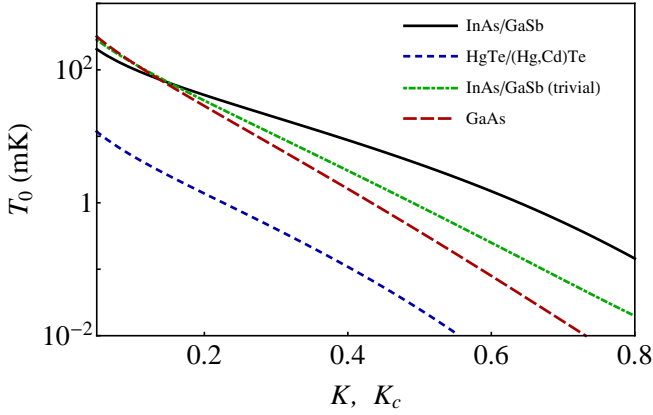}
\caption{Transition temperature $T_{0}$ as a function of the Luttinger liquid parameter for various materials. We take $K$ for 2DTIs and $K_{c}$ (with fixed $K_{s}=1$) for nonhelical channels.
The other parameters are listed in Table~\ref{Tab:parameters}. }
\label{Fig:T0_K}
\end{figure}

To proceed, we take $\langle {\bf \tilde{I}} \rangle_{-}$ [see Eq.~(\ref{Eq:order}), with the minus sign] as the ground state for our spin-wave analysis (see Appendix~\ref{Appendix:spin-wave}), from which we obtain the magnon spectrum in an infinitely long edge,
\begin{align}
E_{\textrm{mag}}(q) 
=\frac{I m_{2k_{F}}}{N_{\perp}} \sqrt{2J^{x}_{2k_{F}} \left( 2J^{x}_{2k_{F}} - J^{x}_{q-2k_{F}} -  J^{x}_{q+2k_{F}} \right)}, 
\label{Eq:omega_mag}
\end{align}
which we plot in Fig.~\ref{Fig:magnon}. There is a zero-energy Goldstone mode at $q=0$ as a consequence of the U(1) rotational symmetry, as discussed. Importantly, however, in a {\it finite-size} system the Goldstone mode is gapped out, and the remaining magnon spectrum is basically dispersionless, as the RKKY resonance dip is very narrow. We thus approximate the magnon energy as $E_{\textrm{mag}}(q) \approx \hbar \omega_{\textrm{mag}} \equiv 2I|J^{x}_{2k_{F}}|m_{2k_{F}}/N_{\perp}$, 
allowing us to analytically compute the temperature dependence of the order parameter (see Appendix~\ref{Appendix:spin-wave}),
\begin{equation}
m_{2k_{F}}(T) = 1- \frac{1}{2} \left(\frac{T}{T_{0}}\right)^{3-2K},~~ \textrm{as }T \apprle T_{0}, 
\label{Eq:Bloch}
\end{equation}
shown in the left inset of Fig.~\ref{Fig:magnon}. Here the transition temperature $T_{0}$ is defined such that $m_{2k_{F}}(T_{0})=1/2$, leading to
\begin{eqnarray}
k_{B}T_{0} &=&  \left[ \frac{A_{0}^2 I^2 }{3N_{\perp}}
\left( \frac{\Delta}{2\pi K} \right)^{1-2K} C(K) \right]^{1/(3-2K)}, 
\label{Eq:T0} \\
C(K)&\equiv& \frac{\sin (\pi K)}{16\pi^3} 
\left| \frac{ \Gamma\left( 1-K\right) \Gamma\left( K/2 \right)} {\Gamma\left[ (2-K)/2 \right]} \right|^{2}, 
\end{eqnarray} 
which depends crucially on the Luttinger liquid parameter $K$, as demonstrated in Fig.~\ref{Fig:T0_K}. As discussed above, in nonhelical wires the effective Luttinger liquid parameter $K_\textrm{wire}$ is bounded by $1/2$, so the feedback effect from the Overhauser field is essential to enhance $T_{0}$ to millikelvin range~\cite{Braunecker:2009a,Braunecker:2009b,Meng:2014a,Hsu:2015}. In contrast, there is no such lower bound for $K$ for a helical edge, and we obtain $T_{0}$ in the order of tens of mK. The helical character of the edge states thus substitutes for the role of the feedback effect on boosting $T_{0}$. Since the fractional power-law dependence of $T_0$ on $N_{\perp}$ gives $T_{0} \propto N_{\perp}^{-0.38}$ (assuming $K=0.2$), increasing $N_{\perp}$ by a factor of 10 only decreases $T_{0}$ by a factor of 2.4, suggesting that a moderate change in $N_{\perp}$ does not lead to a significant change in the estimated value of $T_{0}$.

Due to the temperature dependencies of the RKKY coupling strength [see Eq.~(\ref{Eq:JRKKY})] and of the order parameter [see Eq.~(\ref{Eq:Bloch})], the magnon excitation energy also depends on the temperature. In the right inset of Fig.~\ref{Fig:magnon}, we plot the temperature dependence of the magnon energy, which grows upon decreasing the temperature as a power law, 
\begin{eqnarray}
\hbar \omega_{\textrm{mag}} (T)&=& \frac{2I}{N_{\perp}} |J^{x}_{2k_{F}}(T)|m_{2k_{F}} (T).
\label{Eq:omega_mag_T} 
\end{eqnarray} 
This temperature dependence affects the efficiency of the magnon-mediated backscattering, and therefore enters the magnon-induced resistance [see Eq.~(\ref{Eq:R_mag-em}) below]. In addition, the gap $\Delta_{\textrm{m}}$ opened in the electron spectrum also depends on the temperature,
\begin{eqnarray}
\Delta_{\textrm{m}} (T) &=&  \Delta \left[ \frac{2K A_{0}I m_{2k_{F}}(T)}{\Delta} \right]^{1/(2-K)}.
\label{Eq:Delta_m_T}
\end{eqnarray}
We obtained this formula using a self-consistent variational approach~\cite{Giamarchi:2003}. This gap is, however, below the Fermi surface, and thus not directly observable in transport experiments.
Since the spiral nuclear spin order has no influence on the electron subsystem at the Fermi surface, the previously considered detection methods~\cite{Braunecker:2009a,Braunecker:2009b,Hsu:2015} are not directly applicable. In a clean and short system the edge states remain gapless despite of the formation of the spiral nuclear spin order. Nevertheless, the gap below the Fermi surface provides an alternative to detect the spiral nuclear spin order, which we discuss in the following subsection.

\subsection{Experimental signatures of the nuclear spin order~\label{SubSec:signature}}
 
The gap $\Delta_{\textrm{m}}$ below the Fermi surface results in experimental signatures which can reveal the spiral order.
Since the nuclear spin dynamics  is much slower than the one of electrons, one may change the gate voltage quickly to shift the electron Fermi energy in the gap, while the pitch of the nuclear spin order, and therefore the position of the gap, remains fixed. Then, indirect evidences for the spiral order can be searched for by measuring the dc conductance, which reduces to zero if the Fermi energy is placed inside the gap. In addition to the gap position, this measurement can also determine the position of the charge neutral point, which would otherwise be difficult to locate due to the constant density of states for the edge states. 
Alternatively, one can reach states away from the Fermi surface with finite-frequency measurements. For instance, the Drude peak in the ac conductivity shifts from zero frequency to a finite frequency associated with the gap, when the scanned Fermi energy is inside the gap.

With these intuitions, we now proceed to explicit formulas. We first consider the case when the Fermi energy is outside of the gap. 
In optical experiments, the ac conductivity can be measured without the influence of the leads. Following Ref.~\cite{Giamarchi:2003}, we compute the ac conductivity (see Appendix~\ref{Appendix:transport} for the details),
\begin{eqnarray}
\sigma(\omega) &=& \frac{e^2}{\hbar} (uK) \left[ \delta \left(\omega\right) + \frac{i}{\pi} \mathcal{P}\left(\frac{1}{\omega}\right) \right],
\label{Eq:sigma_nl_omega}
\end{eqnarray}
with the Dirac delta function $\delta (x)$ and the principal value $\mathcal{P}(x)$. The real part of the ac conductivity shows a Drude peak at zero frequency with the weight $ uK (e^2/\hbar)$, and the imaginary part is connected to the real part through the Kramers-Kronig relations.

\begin{figure}[t]
\centering
\includegraphics[width=\linewidth]{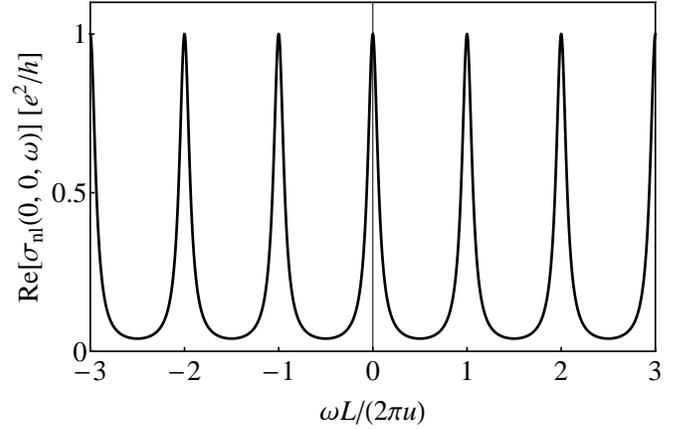}
\caption{The dependence of the real part of the nonlocal conductivity Re[$\sigma_{\textrm{nl}}(0,0,\omega)$] on angular frequency ($\omega$), obtained from Eq.~(\ref{Eq:sigma_nl}). Here we take the Luttinger liquid parameter $K=0.2$ ($K_{\textrm{L}}=1$) for the edge (lead) electrons. }
\label{Fig:sigma}
\end{figure}

When measuring the charge transport through edge states over the finite length $L$, however, the effects of the Fermi liquid leads must be incorporated. To this end, we apply the Maslov-Stone approach~\cite{Maslov:1995,Ponomarenko:1995,Safi:1995,Meng:2014b,Muller:2017} to compute the nonlocal conductivity $\sigma_{\textrm{nl}}$ and the dc conductance $G_{\textrm{dc}}$ by modeling the leads as a Tomonaga-Luttinger liquid with a different parameter $K_{\textrm{L}}$. The nonlocal conductivity $\sigma_{\textrm{nl}}$ relates the charge current $I_{\textrm{c}}$ to the external electric field $E_{\textrm{ext}}$ through
\begin{eqnarray}
I_{\textrm{c}}(r,t) &=& \int_{-L/2}^{L/2} dr' \int \frac{d\omega}{2\pi} \; e^{-i\omega t} \sigma_{\textrm{nl}}(r,r',\omega) E_{\textrm{ext}}(r',\omega). \nonumber \\
\label{Eq:I_sigma_E}
\end{eqnarray} 
In general, the nonlocal conductivity depends on both $r$ and $r'$, and here we give only its expression at the origin [$(r,r')=(0,0)$], which is related to the dc conductance by $G_{\textrm{dc}} = \lim_{\omega \rightarrow 0} \textrm{Re}[\sigma_{\textrm{nl}}(0,0,\omega)] $. With the details of derivation presented in Appendix~\ref{Appendix:transport}, the real and imaginary parts of the nonlocal conductivity at $(r,r')=(0,0)$ are given by
\begin{subequations}
\label{Eq:sigma_nl}
\begin{align}
&\textrm{Re}[\sigma_{\textrm{nl}}(0,0,\omega)] = \frac{e^2}{h} \frac{K^2}{K_{\textrm{L}}} \frac{1}{
\sin^2 \left( \frac{\omega L}{2u} \right) + \left( \frac{K}{K_{\textrm{L}}} \right)^2 \cos^2 \left( \frac{\omega L}{2u} \right) }, \\ 
&\textrm{Im}[\sigma_{\textrm{nl}}(0,0,\omega)] = \frac{e^2 K }{h} \tan \left( \frac{\omega L}{2u} \right) \frac{ \left( \frac{K}{K_{\textrm{L}}} \right)^2 - 1}{
\left( \frac{K}{K_{\textrm{L}}} \right)^2 + \tan^2 \left( \frac{\omega L}{2u} \right) }. 
\end{align}
\end{subequations}
As shown in Fig.~\ref{Fig:sigma}, the real part $\textrm{Re}[\sigma_{\textrm{nl}}(0,0,\omega)]$ oscillates between the maximal value $K_{\textrm{L}} (e^2/h)$ at the angular frequencies $\omega = 2n \pi u/L$ and the minimal value $(K^2/K_{\textrm{L}}) (e^2/h)$ at the angular frequencies $\omega = (2n+1) \pi u/L$ with an integer $n$, similar to a fractional helical Tomonaga-Luttinger liquid~\cite{Meng:2014b}. 
Using Eq.~(\ref{Eq:sigma_nl}), we get the following expression for the dc conductance,
\begin{eqnarray}
G_{\textrm{dc}} &=& \frac{e^2}{h} K_{\textrm{L}}.
\label{Eq:G_lead}
\end{eqnarray}
Importantly, $G_{\textrm{dc}} $ is independent of the Luttinger liquid parameter $K$ of the edge states. We note that Eq.~(\ref{Eq:G_lead}) is valid in a short edge $L\ll \xi_{\textrm{hf}}$, where the resistance caused by the nuclear spins [Eqs.~(\ref{Eq:R_hf_L})--(\ref{Eq:R_hf_gapped})] is insignificant.
Even though Eq.~(\ref{Eq:G_lead}) suggests that measuring $G_{\textrm{dc}}$ in the gapless regime does not reveal any feature of the nuclear spin order or the Tomonaga-Luttinger liquid, it can be used to contrast the measurement when the Fermi energy is in the gap, as we discuss below.

\begin{figure}[t]
\centering
\includegraphics[width=\linewidth]{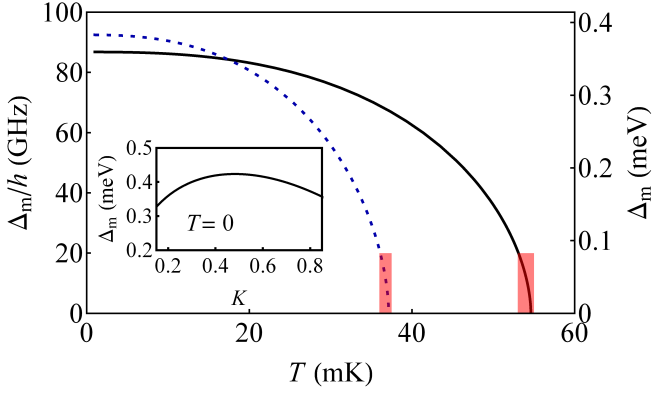}
\caption{The dependence of the gap $\Delta_{\textrm{m}}$ opened in the electron spectrum on temperature $T$. On the left axis, the gap value is converted into the frequency $\Delta_{\textrm{m}}/h$. The Luttinger liquid parameter is taken to be $K=0.2$ ($K=0.25$) for the black solid (blue dashed) curve. The other parameters are listed in Table~\ref{Tab:parameters}. The red shaded region marks the temperature region within which $\Delta_{\textrm{m}}/h \le 20~$GHz. Inset: zero-temperature gap $\Delta_{\textrm{m}}(T=0)$ as a function of $K$.}
\label{Fig:Delta_m_T}
\end{figure}

We now consider the case when the Fermi energy is quickly tuned into the gap $\Delta_{\textrm{m}}$, where the action acquires a sine-Gordon term [see Eq.~(\ref{Eq:S_SG})]. The dc conductance in this case is absent, $G_{\textrm{dc}}=0$, instead of being given by Eq.~(\ref{Eq:G_lead}). Therefore, the zero dc conductance in the range $[-2\hbar v_{F} k_{F} - \Delta_{\textrm{m}}/2, -2\hbar v_{F} k_{F} + \Delta_{\textrm{m}}/2]$ when scanning the Fermi energy by a back gate can serve as an experimental signature for the spiral order, as well as a method to determine the gap value. 

An alternative is provided by the ac conductivity probed optically with the Fermi energy inside the gap,
\begin{eqnarray}
\sigma(\omega) &=& \frac{e^2}{\hbar} (uK) \left[ \delta\left(\omega-\frac{\Delta_{\textrm{m}}^2}{\hbar^2\omega}\right) \right. \nonumber\\
&& \hspace{0.55in} + \left. \frac{i}{\pi} \mathcal{P}\left(\frac{1}{\omega- \Delta_{\textrm{m}}^2/(\hbar^2\omega)}\right) \right]. 
\end{eqnarray}
In this case, the Drude peak is shifted to a finite frequency corresponding to $\Delta_{\textrm{m}}$. As shown in Fig.~\ref{Fig:Delta_m_T}, the gap $\Delta_{\textrm{m}}$ [see Eq.~(\ref{Eq:Delta_m_T})] depends on the temperature, so does the position of the Drude peak. Therefore, tracking the evolution of the Drude peak position with the temperature can then verify the temperature dependence of the gap $\Delta_{\textrm{m}}$, and thus the generalized Bloch law given by Eq.~(\ref{Eq:Bloch}). 
We find that to access the Drude peak, say, at $T=54~$mK, it requires a microwave source with the frequency, $\Delta_{\textrm{m}}(T=54~$\textrm{mK}$)/h \approx 12~$GHz, which is experimentally accessible. 

We conclude this section with some remarks on the proposed microwave measurements. First, the maximal frequency that a microwave source can reach sets a practical constraint for observing the Drude peaks. Thus, in Fig.~\ref{Fig:Delta_m_T}, we mark the temperature region in which the corresponding gap value can be reached by assuming this maximal frequency to be 20~GHz. 
Second, the temperature fluctuations near the transition temperature lead to the fluctuation of the gap as indicated in Fig.~\ref{Fig:Delta_m_T}, so a precise temperature control would be required for a clear peak in the measurements.
Third, it is necessary for the edge length to be longer than the Fabry-Perot length defined as $L_{\textrm{FP}} \equiv h v_{F}/\Delta_{\textrm{m}}$ such that the effect of the leads is negligible. For our parameters, we find $L_{\textrm{FP}}$ to be in the order of $\mu$m.

\subsection{Self-consistency of the RKKY approach~\label{SubSec:consistency}}

In this subsection we comment on the self-consistency condition of the RKKY approach, which allows us to derive the RKKY interaction Eq.~\eqref{Eq:H_RKKY} from the hyperfine interaction Eq.~\eqref{Eq:H_hf_1d}~\cite{Simon:2008}.
Even though similar discussions are also given in Refs.~\cite{Braunecker:2009b,Meng:2014a,Hsu:2015} for nonhelical systems, here we point out the importance of the finite-size effect.

For the self-consistency of the RKKY approach, we examine the following conditions. First, the RKKY approach requires the energy scale of the electron subsystem to be larger than the coupling between the electron and nuclear subsystems, such that the higher-order terms after performing the Schrieffer-Wolff transformation can be dropped. As mentioned in Sec.~\ref{SubSec:RKKY}, this is justified by the weak hyperfine coupling compared to the electron Fermi energy. Second, the separation of the time scales of the electron and nuclear spin dynamics can be examined by comparing the Fermi velocity $v_F$ and the magnon velocity~\cite{Braunecker:2009b}. We check that, around $T_{0}$, $v_F$ is larger than the maximal magnon velocity, computed from the slope of the magnon spectrum in the vicinity of zero momentum [see Fig.~\ref{Fig:magnon}]. This verifies that the dynamics of the nuclear spins is slower than the electrons. Finally, we check that the energy scale of the RKKY term Eq.~\eqref{Eq:H_RKKY} is bounded by the original hyperfine interaction Eq.~\eqref{Eq:H_hf_1d}. 
Since $E_{\rm RKKY} \propto |\tilde{I}_{q=2k_F}|^2$ while $E_{\rm hf} \propto \tilde{I}_{2k_F}$, the ratio $E_{\rm RKKY}/E_{\rm hf} \propto \tilde{I}_{2k_F} \propto m_{2k_F}(T)$ decreases with the fraction of the ordered nuclear spins $m_{2k_F}$. Therefore, $E_{\rm RKKY}$ is bounded by $E_{\rm hf}$ for $T \apprge T_{0}$.
  
At lower temperatures, however, there arise some subtleties when examining the ratio $E_{\rm RKKY}/E_{\rm hf}$.
First, the magnitudes of the RKKY coupling $|J_{2k_F}^{x}|$ and thus $E_{\rm RKKY}$ diverge at zero temperature, whereas $E_{\rm hf}$ does not. As a result, upon decreasing the temperature, the energy scale of $E_{\rm RKKY}$ inevitably exceeds $E_{\rm hf}$ at some point. This issue arises because Eq.~\eqref{Eq:JRKKY} was derived assuming an infinite system~\cite{Giamarchi:2003}, leading to an unphysical divergence at zero temperature. For a realistic system, the finite-size effect has to be taken into account. Since the thermal length $\lambda_T$ below $T_0$ is comparable with a typical edge length of $O(10~\mu{\rm m})$, the finite-size effect becomes relevant at such low temperatures, and the edge length emerges as a cutoff for the divergence in Eq.~\eqref{Eq:JRKKY}. Second, in the ordered phase, the hyperfine coupling in $E_{\rm hf}$ is renormalized by the electron-electron interaction~\cite{Braunecker:2009b,Meng:2014a,Hsu:2015}. Third, the electron subsystem can also be affected by the ordering of the nuclear spins. The direct comparison between $E_{\rm RKKY}$ and $E_{\rm hf}$ then incorrectly neglects the different contributions from the electron subsystem before and after applying the RKKY approach. 

To reflect these issues and make a sensible check on the self-consistency, 
we introduce the edge length $L$ as a cutoff by replacing $q \rightarrow q + i\pi/L$ in the zero-temperature expression of the RKKY coupling, which is given by Eq.~(30) in Ref.~\cite{Braunecker:2009b}. This gives $E_{\rm RKKY} / E_{\rm hf} \sim O(10)$ for $L = 10~\mu$m and $K=0.2$. 
This value is the upper bound for the ratio, and it will be reduced by finite temperature and therefore we conclude that the bound is fulfilled up to a numerical factor of order unity. Using the finite-size expression of the RKKY interaction, we now reestimate the value for $T_0$, which is reduced. Importantly, this reduction is modest, since $T_0$ depends on the magnitude of the RKKY coupling only weakly, as indicated by Eq.~\eqref{Eq:T0}. More precisely, it is given by a fractional power law ($T_0 \propto |J_{2k_F}^{x}|^{0.38}$ for $K=0.2$), and for $L=10$~$\mu$m the reduction is a factor of $\sim 0.4$.
We therefore conclude that at typical parameters that we use, our approach is self-consistent. Due to analytical inconveniences accompanying the finite-size regularization, we use the RKKY interaction of an infinite system, Eq.~\eqref{Eq:JRKKY}, elsewhere in the article. This somewhat (by a factor of order unity) overestimates the transition temperature, an error which is of little importance here.

\section{Resistance in the ordered phase~\label{Sec:ordered}}

\begin{figure}[t]
\centering
\includegraphics[width=0.5\linewidth]{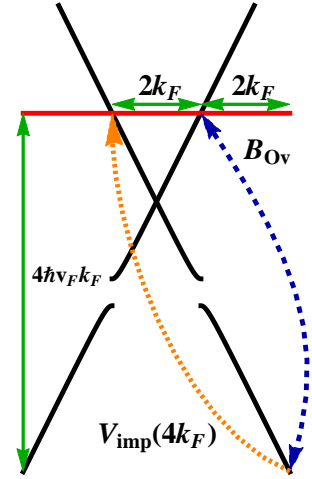}
\caption{Schematics of Overhauser-field-assisted backscattering on impurities. The Overhauser field admixes $R_{\downarrow}(q)$ and $L_{\uparrow}(q+2k_{F})$. 
The mixed $L_{\uparrow}$ component can then forward scatter to the Fermi surface due to the impurity random potential $V_{\textrm{imp}}$, giving rise to the effective backscattering.
For clarity, other scattering processes (e.g. electrons with the opposite velocities) are not shown.}
\label{Fig:backscattering}
\end{figure}

After discussing the RKKY-induced spiral order, we consider how it modifies the transport properties of the edge states. We find that there are two additional backscattering mechanisms in the ordered phase. First, the spin-flip backscattering can arise as a combination of the Overhauser field induced by the nuclear spin order and the impurities. The former provides the spin flip, whereas the $4k_{F}$ component of the random potential of the latter provides the necessary momentum. This Overhauser-field-assisted backscattering process is sketched in Fig.~\ref{Fig:backscattering}. 
Second, magnons, the excitations of the nuclear spin ground state, can also cause electron backscattering. 
It differs from the electron-nuclear spin coupling in the disordered phase, since now it takes a finite exchange energy, set by the magnon energy, for the electron and nuclear spins to flip-flop. Therefore, in the ordered phase it costs energy $\hbar \omega_{\rm mag}$ for the electrons to backscatter.
In addition, thermally excited magnons can be absorbed by electrons and cause additional backscattering. In the following, we show how these backscattering mechanisms arise from the hyperfine interaction $\mathit{H}_{\textrm{hf}}$ defined in Eq.~\eqref{Eq:H_hf}.

To this end, we perform the Holstein-Primakoff transformation [see Appendix~\ref{Appendix:spin-wave} for details] on the hyperfine interaction [see Eq.~\eqref{Eq:H_hf_1d}], which is then written as $\mathit{H}_{\textrm{hf}} = \langle \mathit{H}_{\textrm{hf}} \rangle_{\textrm{gs}} + \mathit{H}_{\textrm{e-mag}}$, where $\langle \cdots \rangle_{\textrm{gs}}$ denotes the expectation value with respect to the nuclear spin ground state~\cite{Holstein:1940}. The first term arises from the ground state of the spiral order,
\begin{eqnarray}
 \langle \mathit{H}_{\textrm{hf}} \rangle_{\textrm{gs}} &=&   \frac{B_{\textrm{Ov}}}{2\pi a} \int dr \; \cos \left[ 2\phi(r) - 4k_{F} r \right],
\label{Eq:H_Ov}
\end{eqnarray} 
with the Overhauser field given by $B_{\textrm{Ov}} \equiv A_{0}I m_{2k_{F}}$.
The second term describes the coupling between the electrons and the magnons, 
\begin{eqnarray}
\mathit{H}_{\textrm{e-mag}} &\approx& 
\frac{ A_{0}  }{2  L^2} \sqrt{ \frac{I m_{2k_{F}}}{2 N_{\perp}}} \sum_{q,q'} \; \frac{1}{i} \left( b_{q'}^{\dagger} + b_{- q'} \right)  \nonumber \\
&& \hspace{0.5in} \times L_{\uparrow}^{\dagger}(q) R_{\downarrow}(q+q'-2k_{F}) + \textrm{H.c.},
\label{Eq:H_e-mag_FT} 
\end{eqnarray}
where we keep the lowest-order backscattering terms in magnon operators.
In the above, $b_{q}^{\dagger}$ ($b_{q}$) creates (annihilates) a magnon with momentum $q$. 
In the following subsections, we then investigate the edge resistance caused by these additional backscattering processes defined by Eqs.~(\ref{Eq:H_Ov}) and (\ref{Eq:H_e-mag_FT}).

\subsection{Impurity-induced resistance in the ordered phase~\label{SubSec:hx-bs}}

The Overhauser field [see Eq.~(\ref{Eq:H_Ov})] contains an oscillating integrand except for the special case $4k_{F}a = \textrm{integer} \times \pi$, which we do not consider. Therefore, Eq.~(\ref{Eq:H_Ov}) is irrelevant in the RG sense, and does not cause any electron backscattering at the Fermi surface on its own. Nonetheless, it causes a mixing of the right- and left-moving electrons with opposite spins, lifting the topological protection of the helical edge states against impurities. To proceed, we model the impurity Hamiltonian as 
\begin{eqnarray}
\mathit{H}_{\textrm{imp}} &=& \int dr \, V_{\textrm{imp}}(r) \left[R_{\downarrow}^{\dagger}(r) R_{\downarrow}(r) + L_{\uparrow}^{\dagger}(r) L_{\uparrow}(r) \right] ,
\label{Eq:H_imp}
\end{eqnarray} 
where the Gaussian random potential $V_{\textrm{imp}}(r)$ satisfies $\overline{V_{\textrm{imp}}(r) V_{\textrm{imp}}(r')} = M_{\textrm{imp}} \delta (r-r')$, with $\overline{\cdots}$ denoting the average over the random potential. We estimate the impurity strength $M_{\textrm{imp}}=\hbar^2 v_{F}^2/(2\pi \lambda_{\textrm{mfp}})$ with the mean free path of the 2DTI bulk of $\lambda_{\textrm{mfp}} \sim 0.1\text{--}1~\mu$m~\cite{Konig:2007,Li:2015}, as listed in Table~\ref{Tab:parameters}.
To derive the effective action for the nuclear-order-assisted backscattering on impurities, we perform a Schrieffer-Wolff transformation~\cite{Schrieffer:1966,Bravyi:2011} and average over impurities~\cite{Giamarchi:2003}. We defer the details of calculations in Appendix~\ref{Appendix:SW}. Here, we present the result,
\begin{eqnarray}
\frac{\delta \mathit{S}_{\textrm{hx}}}{\hbar} 
&=& - \frac{M_{\textrm{hx}}}{(2\pi \hbar a)^2  }  \int_{u|\tau-\tau'|>a}  dr d\tau d\tau' \; \nonumber \\
&& \hspace{0.8in} \times \cos \left[ 2 \phi (r,\tau)- 2\phi (r,\tau') \right],
\label{Eq:S_hx-bs} 
\end{eqnarray}
which is identical to Eq.~(\ref{Eq:S_hf-bs}) upon replacing the coupling 
$M_{\textrm{hf}} \to M_{\textrm{hx}} \equiv M_{\textrm{imp}}B_{\textrm{Ov}}^2/(64 \hbar^2 v_{F}^2 k_{F}^2)$.
Therefore, the RG flow equations can be derived as in Appendix~\ref{Appendix:RG}, leading to a set of RG flow equations identical to Eq.~(\ref{Eq:RG_hf}) with the replacement $D_{\textrm{hf}} \rightarrow D_{\textrm{hx}} \equiv 2a M_{\textrm{hx}} /(\pi \hbar^{2} u^{2})$.
We then find that Eq.~(\ref{Eq:S_hx-bs}) is RG relevant for $K(l)<3/2$, which leads to the Anderson-type localization in an edge longer than the associated localization length,
\begin{equation}
\xi_{\textrm{hx}} = a D_{\textrm{hx}}^{-1/(3-2K)}, 
\end{equation}
depending on the temperature through $m_{2k_{F}}(T)$ [see Eq.~(\ref{Eq:Bloch})].
Since for the above values of the mean free path this backscattering strength is comparable to the strength of backscattering on disordered nuclear spins, the  localization length at zero temperature $\xi_{\textrm{hx}}(T=0)$ is also comparable to $\xi_{\textrm{hf}}$. We define the characteristic temperature $T_{\textrm{hx}} \equiv \hbar u / [k_{B} \xi_{\textrm{hx}}(T=0)]$ through the zero-temperature localization length, and find that typically $T_{\textrm{hx}} >T_{0}$. This means for a sufficiently long edge $L > \xi_{\textrm{hf}} \approx \xi_{\textrm{hx}}(T=0)$, the electrons get localized by the impurities once the nuclear spins start to order at $T_{0}$.

\begin{figure}[t]
\centering
\includegraphics[width=\linewidth]{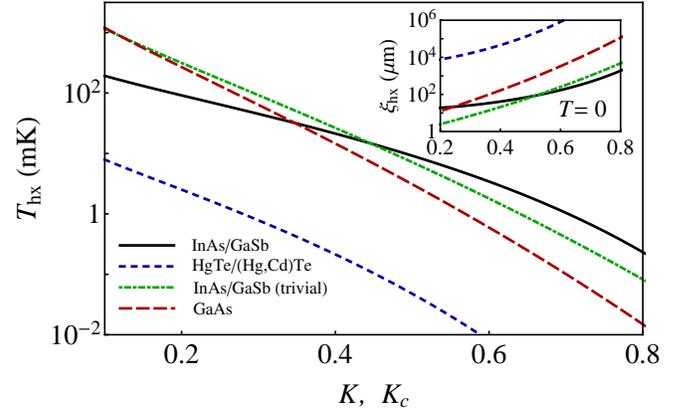}
\caption{Characteristic temperature $T_{\textrm{hx}}$ for various materials as a function of the Luttinger liquid parameter  [$K$ for 2DTI and $K_{c}$ (with fixed $K_{s}=1$) for nonhelical channels]. We take $\lambda_{\textrm{mfp}} = 1~\mu$m, and the other parameters are listed in Table~\ref{Tab:parameters}. Inset: the zero-temperature localization length $\xi_{\textrm{hx}}(T=0)$ as a function of the same variable.}
\label{Fig:T_hx_K}
\end{figure}

In Fig.~\ref{Fig:T_hx_K}, we plot $T_{\textrm{hx}}$ and $\xi_{\textrm{hx}}(T=0)$ as functions of the Luttinger liquid parameter for various materials. Again, a common property shared by all the curves is that the backscattering effects are enhanced by the electron-electron interaction. In contrast to the disordered phase (see Fig.~\ref{Fig:T_hf_K}), however, here the estimated quantities for the helical and nonhelical states are comparable in the strongly interacting regime, indicating that the localization effects in the ordered phase are not as markedly different for a helical and a nonhelical channel as in the disordered phase. The reason behind this is that the Overhauser field in a spin-degenerate wire provides a synthetic spin-orbit interaction in the ordered phase, making the remaining gapless electrons helical~\cite{Braunecker:2009a,Braunecker:2009b,Braunecker:2010,Klinovaja:2013b,Braunecker:2013,Hsu:2015}. After ordering, the effective Luttinger liquid parameter of the remaining gapless modes is not bounded by $1/2$ anymore. As a consequence, the estimated values of $T_{\textrm{hx}}$ and $\xi_{\textrm{hx}}(T=0)$ in the helical and nonhelical systems become similar in the presence of strong electron-electron interaction.

Since typically $T_{\textrm{hx}} > T_{0}$, the impurities induce an exponentially growing resistance below $T_{0}$, 
\begin{eqnarray}
R_{\textrm{hx}} (T) &\propto& R_{0} \frac{\pi D_{\textrm{hx}}L }{2K^2 a}  e^{\Delta_{\textrm{hx}} /(k_{B}T)},
\label{Eq:R_hx}
\end{eqnarray}
with a gap $\Delta_{\textrm{hx}} = \Delta \left( 2 K D_{\textrm{hx}} \right)^{1/(3-2K)}$. Despite its similarity to Eq.~\eqref{Eq:R_hf_gapped}, $R_{\textrm{hx}}$ contains the temperature-dependent $D_{\textrm{hx}}$ and $\Delta_{\textrm{hx}}$ while their counterparts $D_{\textrm{hf}}$ and $\Delta_{\textrm{hf}}$ are independent of $T$. As a result, the resistances arising from the two scenarios [Eq.~\eqref{Eq:R_hf_gapped} versus Eq.(\ref{Eq:R_hx})] are distinct due to the different temperature dependencies of the gaps and prefactors, as well as the dependence of $D_{\textrm{hx}}$ and $\Delta_{\textrm{hx}}$ on $\epsilon_{F}$ and $M_{\textrm{imp}}$.

Before moving to the magnon-mediated backscattering, let us comment on two complications not considered in this work. First, we remark that an applied voltage may lead to the nuclear spin polarization along the spin quantization ($z$) axis~\cite{Lunde:2012,Kornich:2015,Russo:2017}, and thus modify the nuclear spin order. While the $z$ component of the nuclear spin polarization does not directly cause the spin-flip backscattering, it reduces the $xy$ components of the Overhauser field from $B_{\textrm{Ov}}$ to $B_{\textrm{Ov}} \sqrt{1-P_{N}^{2}(T)}$, with a temperature-dependent factor, $P_{N}(T)$. However, unless the nuclear spins are nearly full-polarized, which requires a very high applied voltage at very low temperatures, the residual $xy$ components of the Overhauser field can still cause the spin-flip backscattering on impurities. Therefore, the voltage-induced dynamic nuclear polarization would not alter our conclusion qualitatively.

Second, in the ordered phase the gap $\Delta_{\textrm{hx}}$ reduces the RKKY coupling, and therefore the strength of the nuclear-order-assisted backscattering. However, since the effective range of the RKKY coupling is related to the electron Fermi wavelength $\lambda_{F} \equiv 2\pi / k_{F}$, the gapped electrons can still mediate the RKKY interaction within the scale of $\lambda_{F}$, provided that it is much shorter than the length scale associated with the gap, $\hbar v_{F} / \Delta_{\textrm{hx}}$, as discussed in Refs.~\cite{Klinovaja:2013b,Meng:2014a,Hsu:2015}. We thus expect our results to remain qualitatively valid if the condition $\lambda_{F} \ll \hbar v_{F} /\Delta_{\textrm{hx}}$ holds.
We have checked that for our case it is fulfilled, so that the RKKY interaction remains effective, even though the coupling strength is reduced by the gap $\Delta_{\textrm{hx}}$.

\subsection{Resistance due to the magnon-mediated backscattering~\label{SubSec:mag-bs}}

We now turn to the electron-magnon interaction described by Eq.~\eqref{Eq:H_e-mag_FT} with the magnon dispersion given by Eq.~(\ref{Eq:omega_mag}). With the approximated magnon energy [see Eq.~(\ref{Eq:omega_mag_T})], we are able to reformulate the electron-magnon backscattering as an electron-phonon backscattering problem. In particular, it is analogous to a Tomonaga-Luttinger liquid consisting of spinless fermions coupled to dispersionless phonons~\cite{Voit:1987,Voit:1995}.
We then proceed by integrating out the magnons, and obtain the contribution to the effective action from the magnon-mediated backscattering. The details of calculations are relegated to Appendix~\ref{Appendix:mag-bs}, in which we get the following expressions,
\begin{subequations}
\label{Eq:S_mag-bs}
\begin{align}
&\delta \mathit{S}_{\textrm{mag}}= \delta \mathit{S}_{\textrm{mag}}^{\textrm{em}} + \delta \mathit{S}_{\textrm{mag}}^{\textrm{abs}},  \\
&\frac{\delta \mathit{S}_{\textrm{mag}}^{\textrm{em}}}{\hbar} = - \frac{M_{\textrm{mag}}}{(2\pi \hbar a )^2  }  \int_{u|\tau-\tau'|>a} dr d\tau d\tau' \; e^{-\omega_{\textrm{mag}} |\tau-\tau'| } \nonumber \\
&  \hspace{0.1in}  \times
\left[  1 + n_{B}(\hbar \omega_{\textrm{mag}}) \right] \cos \left[  2 \phi (r,\tau)- 2\phi (r,\tau') \right] ,
 \label{Eq:S_mag-em} \\
&\frac{\delta \mathit{S}_{\textrm{mag}}^{\textrm{abs}}}{\hbar} = - \frac{M_{\textrm{mag}}}{(2\pi \hbar a )^2  }  \int_{u|\tau-\tau'|>a} dr d\tau d\tau' \; e^{\omega_{\textrm{mag}} |\tau-\tau'| } \nonumber \\
&  \hspace{0.1in}  \times
 n_{B}(\hbar \omega_{\textrm{mag}}) \cos \left[  2 \phi (r,\tau)- 2\phi (r,\tau') \right],
\label{Eq:S_mag-abs}
\end{align}
\end{subequations}
with $M_{\textrm{mag}} \equiv A_{0}^2 a I/(4N_{\perp})$ being the backscattering strength and the Bose-Einstein distribution given by
\begin{eqnarray}
n_{B}(E)  &=& \frac{1}{e^{E / (k_{B}T)} - 1}.
\end{eqnarray}
In comparison with Eqs.~(\ref{Eq:S_hf-bs}) and (\ref{Eq:S_hx-bs}), the magnon-mediated backscattering acquires extra exponential factors $e^{\pm \omega_{\textrm{mag}} |\tau-\tau'|}$ in the integrand, corresponding to the process where a magnon is absorbed ($+$) or emitted ($-$) because of a finite energy exchange due to the electron spin flip through a magnon. In addition, the efficiency of the magnon-mediated backscattering depends on the magnon occupation $n_{B}(\hbar \omega_{\textrm{mag}})$, and therefore on the temperature.

One can visualize the effects of the magnon-mediated backscattering on the resistance by examining Eq.~(\ref{Eq:S_mag-bs}): if the magnon energy is much larger than the temperature, the backscattering is suppressed exponentially by either the exponential factor in Eq.~\eqref{Eq:S_mag-em} (for magnon emission), or by the Boltzmann factor of the magnon occupation in Eq.~(\ref{Eq:S_mag-abs}) (for magnon absorption). We then expect that the magnon-induced resistance to be suppressed in the $T \rightarrow 0$ limit. On the other hand, if the magnon energy is comparable to the temperature (that is, when $T$ is near $T_{0}$), the order parameter is small and there are many thermally excited magnons. Then, the electron-magnon backscattering events become efficient and give rise to a resistance similar to the one caused by disordered nuclear spins [Eqs.~\eqref{Eq:R_hf_L}-\eqref{Eq:R_hf_gapped}], which can be considered as the ordered nuclear spins with zero-energy magnons. 

We confirm these observations by performing the RG analysis. In the low-temperature limit, where the contribution from magnon emission dominates, we have $\delta \mathit{S}_{\textrm{mag}} \approx \delta \mathit{S}_{\textrm{mag}}^{\textrm{em}}$ with $n_{B}(\hbar \omega_{\textrm{mag}}) \rightarrow 0$. When the short-distance cutoff $a$ increases under the RG flow, the exponential factor in Eq.~(\ref{Eq:S_mag-em}) decreases, and $\delta \mathit{S}_{\textrm{mag}}^{\textrm{em}}$ becomes vanishingly small as $a>u/\omega_{\textrm{mag}}$. Therefore, we derive the RG flow equations up to the scale $l_{\textrm{mag}}^{*} \equiv \ln [u/(a \omega_{\textrm{mag}})]$ as in Refs.~\cite{Voit:1987,Voit:1995} (similar to the procedure given in Appendix~\ref{Appendix:RG}), which are given by 
\begin{subequations}
\label{Eq:RG_mag-bs}
\begin{align}
&\frac{d Y_{\textrm{mag}} (l) }{d l } =\left[3-2K(l) \right] Y_{\textrm{mag}} (l), \\
&\frac{d K (l) }{d l } = - \frac{K^2 (l) }{2} Y_{\textrm{mag}}(l) \frac{\omega_{\textrm{mag}} a }{u} e^{- \omega_{\textrm{mag}} a(l)/u}, \\
&\frac{du(l)}{dl} = -\frac{u(l)K(l)}{2} Y_{\textrm{mag}}(l) \frac{\omega_{\textrm{mag}} a }{u} e^{-\omega_{\textrm{mag}} a(l)/u}, 
\end{align} 
\end{subequations}
with $Y_{\textrm{mag}} \equiv 2 M_{\textrm{mag}} / (\pi \hbar^2 u \omega_{\textrm{mag}})$.
Using Eq.~(\ref{Eq:RG_mag-bs}), we calculate the resistance caused by the magnon emission at low temperatures $T<T_{x}$. The temperature $T_x$, defined by $\omega_{\textrm{mag}} (T_{x}) = D_{\textrm{mag}}^{1/(4-2K)} u/a$, gives the limit at which the RG flow reaches the strong-coupling regime.
For $T<T_{x}$, we integrate the RG flow up to $l_{\textrm{mag}}^{*} $ to obtain 
\begin{eqnarray}
R_{\textrm{mag}}^{\textrm{em}} (T) &\propto& R_{0} \frac{\pi D_{\textrm{mag}}L}{2 K^2 a} \left[ \frac{K \hbar \omega_{\textrm{mag}}(T) }{\Delta} \right]^{2K-3},
\label{Eq:R_mag-em}
\end{eqnarray} 
with $D_{\textrm{mag}} \equiv 2a M_{\textrm{mag}} /(\pi \hbar^{2} u^{2})$. Upon decreasing the temperature, the resistance due to magnon emission decreases as a power law of the magnon energy, whose temperature dependence is given by Eq.~(\ref{Eq:omega_mag_T}). 
 
\begin{figure}[t]
\centering
\includegraphics[width=\linewidth]{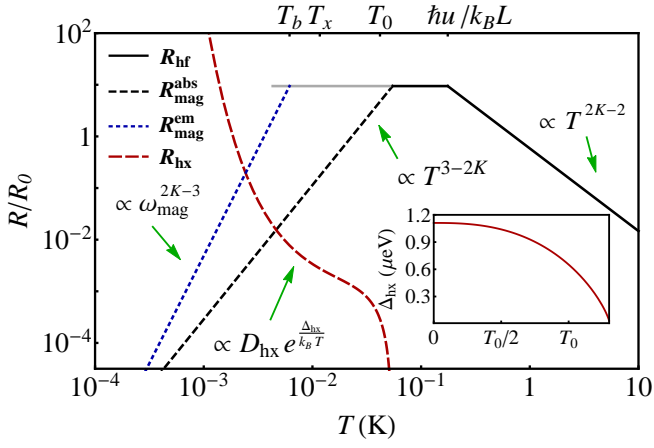}
\caption{A summary of the resistance ($R$) as a function of the temperature $T$ in both the disordered and ordered phases for the parameters of InAs/GaSb. 
We take $K=0.2$, the mean free path $\lambda_{\textrm{mfp}}= 0.1~\mu$m, and $L=10~\mu\textrm{m}$, so that $\xi_{\textrm{hx}} < L < \xi_{\textrm{hf}}$. 
The other parameters are listed in Table~\ref{Tab:parameters}. 
In the disordered phase ($T>T_{0}$), $R$ is given by Eqs.~(\ref{Eq:R_hf_T}) and (\ref{Eq:R_hf_L}) for $T>\hbar u / (k_{B}L) $ and $T<\hbar u / (k_{B}L) $, respectively. 
In the ordered phase ($T<T_{0}$), three backscattering processes contribute to $R$, including the Overhauser-field-assisted backscattering on impurities $R_{\textrm{hx}}$ [Eq.~(\ref{Eq:R_hx})], the magnon emission $R_{\textrm{mag}}^{\textrm{em}}$ [Eq.~(\ref{Eq:R_mag-em})], and the magnon absorption $R_{\textrm{mag}}^{\textrm{abs}}$ [Eq.~(\ref{Eq:R_mag-abs})]. The gray curve gives the upper limit on the sum of the last two.
Inset: $T$ dependence of the gap $\Delta_{\textrm{hx}}$.
}
\label{Fig:R_T_short}
\end{figure}

On the other hand, in the range $T_{x} \apprle T \apprle T_{0}$, the backscattering due to the magnon absorption process Eq.~(\ref{Eq:S_mag-abs}) becomes efficient. In this case, the magnon energy is so low compared to the temperature that it can be approximated by zero. We therefore expect that the associated resistance $R_{\textrm{mag}}^{\textrm{abs}}(T)$ takes the form of $R_{\textrm{hf}}$ [Eqs.~(\ref{Eq:R_hf_L})--(\ref{Eq:R_hf_gapped})], with the strength weighted by the fraction of the disordered nuclear spins. Since typically $T_{0}<T_{\textrm{hf}}$, the regime corresponding to Eq.~(\ref{Eq:R_hf_T}), which is valid only for $T>T_{\textrm{hf}}$, is never reached in the ordered phase.  In addition, we restrict ourselves in the low-bias regime, in which the high-bias resistance Eq.~(\ref{Eq:R_hf_V}) is not relevant.
Both of the remaining equations [see Eqs.~(\ref{Eq:R_hf_L}) and (\ref{Eq:R_hf_gapped})] then give the same temperature dependence,
\begin{eqnarray}
R_{\textrm{mag}}^{\textrm{abs}} (T) &\propto& R_{0} \frac{\pi D_{\textrm{hf}}L}{2 K^2 a} \left[ 1-m_{2k_{F}} (T) \right] , 
\label{Eq:R_mag-abs}
\end{eqnarray} 
decaying as a $T^{3-2K}$. 
In addition to the limit set by $T_{x}$, we note another limit described by $R_{\textrm{mag}} (T) \equiv R_{\textrm{mag}}^{\textrm{em}} (T) + R_{\textrm{mag}}^{\textrm{abs}} (T) \le R_{\textrm{hf}} (T) $, with $R_{\textrm{hf}}(T)$ determined by Eqs.~(\ref{Eq:R_hf_L}) and (\ref{Eq:R_hf_gapped}). This arises from the self-consistency check: 
the resistance from the backscattering that requires an energy [see Eq.~(\ref{Eq:S_mag-bs})] should be bounded by the resistance when such an energy cost is absent [see Eq.~\eqref{Eq:S_hf-bs}]. This allows us to define $T_{b}$ through $R_{\textrm{mag}} (T_{b}) = R_{\textrm{hf}}(T_{b})$, and numerically find $T_{b} \approx T_{x}$ (see Fig.~\ref{Fig:R_T_short}), consistent with the limit set by $T_{x}$. Overall, we find the magnon-induced resistance to be dominated by Eq.~(\ref{Eq:R_mag-em}) for $T \apprle T_{x} \approx T_{b} $, whereas Eq.~(\ref{Eq:R_mag-abs}) also contributes in the range $T_{x} \approx T_{b} \apprle T \apprle T_{0}$.

\section{Discussion~\label{Sec:discussion}}

We now summarize our results on the resistance discussed in Secs.~\ref{Sec:hf-bs} and \ref{Sec:ordered}. For the demonstration, we plot the temperature dependence of the edge resistance of the 2DTI of length $L<\xi_{\textrm{hf}}$ in Fig.~\ref{Fig:R_T_short}. The opposite regime, $L>\xi_{\textrm{hf}}$, has been presented in Ref.~\cite{Hsu:2017}. Well above $T_0$, the resistance $R_{\textrm{hf}}$ (the black solid curve) initially increases as a power law, which becomes a plateau in the range $T_{0}<T<\hbar u / (k_{B}L)$ [Eq.~(\ref{Eq:R_hf_L})]. 
Below $T_0$, at which the nuclear spins form an order, the resistance is initially dominated by the magnon-mediated backscattering. The resistance $R_{\textrm{mag}}^{\textrm{abs}}$ from the magnon absorption drops as a power law (the black dashed curve). Below $T_b \approx T_x$, the resistance due to the magnon emission is given by $R_{\textrm{mag}}^{\textrm{em}}$, which decays as a power law (the blue curve) different from $R_{\textrm{mag}}^{\textrm{abs}}$. Both $R_{\textrm{mag}}^{\textrm{em}}$ and $R_{\textrm{mag}}^{\textrm{abs}}$ decay to zero as $T \rightarrow 0$, as expected. 
At very low temperatures, the established Overhauser field dominates the resistance by allowing backscattering on impurities, leading to an exponential form of the resistance $R_{\textrm{hx}}$ (the red curve).
In addition, in the inset of Fig.~\ref{Fig:R_T_short}, we plot the temperature dependence of the gap $\Delta_\textrm{hx}$, which, as opposed to the constant $\Delta_\textrm{hf}$, distinguishes the two exponential regions due to disordered and ordered nuclear spins as $T>T_{0}$ and $T<T_{0}$, respectively. The temperature dependent gap $\Delta_\textrm{hx}$ can therefore serve as an experimental signature of the spiral nuclear spin order, in addition to those discussed in Sec.~\ref{SubSec:signature}.
We conclude that the nuclear spins, whether ordered or not, suppress the edge conductance of a sufficiently long 2DTI sample as the temperature approaches zero.

Finally, we remark that our estimation with realistic material parameters allows us to discuss the relevance of nuclear spins to the edge resistances of HgTe/(Hg,Cd)Te and InAs/GaSb 2DTIs observed in experiments. As mentioned in Sec.~\ref{Sec:hf-bs}, the effects of nuclear spins in HgTe/(Hg,Cd)Te are insignificant even for strong interactions, suggesting that the observed finite edge resistance in HgTe/(Hg,Cd)Te 2DTIs~\cite{Konig:2007,Gusev:2013,Gusev:2014,Olshanetsky:2015} is unlikely due to nuclear spins.

On the other hand, whether nuclear spins in InAs/GaSb 2DTIs can lead to an appreciable edge resistance depends on the experimental conditions. To be explicit, we expect our mechanism to be relevant for strong interactions, long edge lengths, and low temperatures. Since, however, the interaction parameter $K$ is typically unknown in real samples, it is difficult to draw conclusions on the relevance of the nuclear spins. Recently, a value of $K=0.21$--$0.22$ was extracted in InAs/GaSb 2DTIs~\cite{Li:2015}.~\footnotemark[5] 
However, based on our estimation in Secs.~\ref{Sec:hf-bs} and \ref{Sec:ordered}, the corresponding localization length is much longer than the edge length $L\sim 1~\mu$m of the samples in Ref.~\cite{Li:2015}. We therefore believe that the observed edge resistance in their experiment is probably dominated by other sources than nuclear spins. Nevertheless, we expect the nuclear spins in InAs/GaSb to become relevant for longer samples with small $K$ at low temperatures.

\acknowledgments

We thank P.~P.~Aseev, Y.-Z.~Chou, R.~S.~Deacon, M.~R.~Delbecq, S.~Hoffman, P.~Marra, and K.~Muraki for helpful discussions. This work was supported financially by the the JSPS Kakenhi Grant No. 16H02204, the Swiss National Science Foundation (Switzerland), and the NCCR QSIT.

\appendix

\section{Derivation of the RG flow equations~\label{Appendix:RG}}

In this appendix, we sketch the derivation of the RG flow equations for the backscattering action $\mathit{S}_{\textrm{el}}+\delta \mathit{S}_{\textrm{hf}}$ in the disordered phase. The other backscattering processes [Eqs.~(\ref{Eq:S_hx-bs}) and (\ref{Eq:S_mag-bs})] can be treated similarly. To this end, we compute the correlation function as in Refs.~\cite{Giamarchi:1988,Giamarchi:2003},
\begin{align}
 & \left< e^{i [\phi({\bf r_{1}}) - \phi({\bf r_{2}})] } \right>_{\mathit{S}_{\textrm{el}}+\delta \mathit{S}_{\textrm{hf}}} \nonumber\\
&\hspace{30pt}\equiv \mathit{Z}^{-1} \int D\phi \; e^{-\mathit{S}_{\textrm{el}}/\hbar} e^{-\delta \mathit{S}_{\textrm{hf}}/\hbar}  e^{i [\phi({\bf r_{1}}) - \phi({\bf r_{2}})] },
\end{align}
with the bold font denoting the two-dimensional vectors ${\bf r_{j}} \equiv (r_{j}, y_{j}) = (r_{j}, u \tau_{j})$ and the partition function,
\begin{eqnarray}
\mathit{Z} \equiv \int D\phi \; e^{-(\mathit{S}_{\textrm{el}}+ \delta \mathit{S}_{\textrm{hf}})/\hbar}.
\end{eqnarray} 
We expand the correlation function up to the first-order terms in $D_{\textrm{hf}}$, corresponding to the second-order terms in the random potential $V_{\textrm{hf},2k_{F}}$. The zeroth-order term gives
\begin{eqnarray}
\left< e^{i [\phi({\bf r_{1}}) - \phi({\bf r_{2}})]} \right>_{\mathit{S}_{\textrm{el}}} &=& 
e^{- K F({\bf r_{1}} - {\bf r_{2}}) / 2},
\end{eqnarray}
where we define the function,
\begin{align}
& F({\bf r_{1}} - {\bf r_{2}})\label{Eq:F(r)}\\
&\hspace{20pt}\equiv \frac{1}{2} \ln \left[ \frac{(r_{1}-r_{2})^2 + u^2 (\tau_{1} - \tau_{2})^2 }{a^2} \right] + \frac{t_{\perp}}{K} \cos (2 \theta_{{\bf r_{1}} - {\bf r_{2}}} ), \nonumber
\end{align}
where $\theta_{{\bf r}}$ is the angle between the vector ${\bf r}=(r,y)=(r,u\tau)$ and the spatial coordinate axis $r$, and the $t_{\perp}$ term generated by the RG flow gives the anisotropy between the spatial and the temporal coordinates. The first-order term in $D_{\textrm{hf}}$ is given by 
\begin{eqnarray}
&& \frac{D_{\textrm{hf}}}{ 8\pi a^3 } \int_{|y-y'|>a} dr dy dy' 
\;  \\
&& \hspace{0.2in}  \times 
\left\{ \left< e^{i [\phi({\bf r_{1}}) - \phi({\bf r_{2}})]}  
 \cos \left[ 2\phi (r,\tau) - 2\phi (r,\tau')\right] \right>_{\mathit{S}_{\textrm{el}}} \right. \nonumber\\
&& \hspace{0.25in} \left. - \left< e^{i [\phi({\bf r_{1}}) - \phi({\bf r_{2}})]}  \right>_{\mathit{S}_{\textrm{el}}}
\left< \cos \left[ 2\phi (r,\tau) - 2\phi (r,\tau')\right] \right>_{\mathit{S}_{\textrm{el}}} \right\}.\nonumber
\end{eqnarray}
The correlation function can then be computed along the line of Refs.~\cite{Giamarchi:1988,Giamarchi:2003}, giving Exp$[-K_{\textrm{eff}} F_{\textrm{eff}}({\bf r_{1}} - {\bf r_{2}}) /2]$, where $F_{\textrm{eff}}({\bf r_{1}} - {\bf r_{2}})$ takes the form of Eq.~(\ref{Eq:F(r)}) with the effective parameters,
\begin{subequations}
\begin{eqnarray}
K_{\textrm{eff}} &=& K - \frac{K^2 D_{\textrm{hf}} }{2} \int_{a}^{\infty} \frac{dz}{a} \left( \frac{z}{a}  \right)^{2-2K}  , \\
t_{\perp,\textrm{eff}} &=&  t_{\perp} + \frac{K^2 D_{\textrm{hf}} }{4} \int_{a}^{\infty} \frac{dz}{a} \left( \frac{z}{a}  \right)^{2-2K}.
\end{eqnarray}
\end{subequations}
The RG flow equations can be obtained by increasing the cutoff $a \rightarrow a e^{d l} = a + da $ while keeping the correlation function the same. Finally, we obtain a set of three equations,
\begin{subequations}
\begin{eqnarray}
\frac{d D_{\textrm{hf}} (l) }{d l } &=& \left[3-2K(l)\right] D_{\textrm{hf}} (l), \\
\frac{d K (l) }{d l } &=& - \frac{K^2 (l) }{2} D_{\textrm{hf}}(l), \\
\frac{d t_{\perp} (l)}{d l } &=& \frac{K^2 (l) }{4} D_{\textrm{hf}}(l).
\end{eqnarray}
\end{subequations}
In addition, from Eq.~(\ref{Eq:F(r)}) we see that the renormalization of $t_{\perp}$ is equivalent to that of $u$, leading to 
\begin{eqnarray}
 \frac{du(l)}{dl} &=& -\frac{2 u(l)}{K(l)} \frac{d t_{\perp} (l)}{d l }.
\end{eqnarray}
The RG flow equations are then given in Eq.~(\ref{Eq:RG_hf}) in Sec.~\ref{Sec:hf-bs}. Since the RG flow equations are obtained with the perturbation in $D_{\textrm{hf}}(l)$, they are valid only below the length scale $l_{\textrm{hf}}^{*}$, at which $D_{\textrm{hf}}(l_{\textrm{hf}}^{*}) \sim 1$.

\section{Spin-wave analysis~\label{Appendix:spin-wave}}

In this Appendix, we provide the details of the spin-wave analysis. For the sake of convenience, we start by locally rotating the spin axes, $(\tilde{I}_{j}^{x},\tilde{I}_{j}^{y},\tilde{I}_{j}^{z})$ $\rightarrow$ $(\tilde{I}_{j}^{1},\tilde{I}_{j}^{2},\tilde{I}_{j}^{3})$ such that in the new basis $(\hat{e}_{j}^{1},\hat{e}_{j}^{2},\hat{e}_{j}^{3})$ the ground state of the nuclear spins is described as a uniform ferromagnet, i.e. $\left< \tilde{{\bf I}}(r_{j}) \right> = N_{\perp} I m_{2k_{F}}\, \hat{e}_{j}^{1}$~\cite{Simon:2008,Braunecker:2009b}. In addition, the electron spin operators are also rotated accordingly. To proceed, we perform the Holstein-Primakoff transformation~\cite{Holstein:1940}, in which the nuclear spin operators are written in terms of the ground state of the spiral order in addition to the deviation caused by the magnon excitation, 
\begin{subequations}
\label{Eq:HP_xfmn}
\begin{eqnarray}
\tilde{I}^{1}_{j} &=& N_{\perp}I  - a_{j}^{\dagger} a_{j}, \\
\tilde{I}^{2}_{j} &\approx& \sqrt{\frac{N_{\perp}I }{2}} \left( a_{j} + a_{j}^{\dagger} \right), \\
\tilde{I}^{3}_{j} &\approx& \sqrt{\frac{N_{\perp}I }{2}} \frac{1}{i} \left( a_{j} - a_{j}^{\dagger} \right), 
\end{eqnarray}
\end{subequations}
where $a_{j}(a_{j}^{\dagger})$ is the annihilation (creation) operator of a magnon at site $j$. Note that in this Appendix, we assume $T=0$ for the ease of notation, so that $m_{2k_{F}}=1$. For finite temperatures, where the nuclear spins are partially ordered, the formulas are valid with the replacement, $I \rightarrow I m_{2k_{F}}$. Using Eq.~(\ref{Eq:HP_xfmn}), we derive the magnon Hamiltonian $\mathit{H}_{\textrm{mag}}$ in momentum space from the RKKY interaction given by Eq.~(\ref{Eq:H_RKKY}),
\begin{eqnarray}
\mathit{H}_{\textrm{mag}}
&=& \frac{1}{L} \sum_{q} 
\left( a^{\dagger}_{q}, a_{q} \right)
\left( 
\begin{array}{cc}
h_{11}(q) & h_{12}(q) \\
h_{12}(q) & h_{11}(q) 
\end{array}
\right)
\left(
\begin{array}{c}
a_{q} \\  a^{\dagger}_{q}
\end{array}
\right),\hspace{10pt}
\label{Eq:H_mag1}
\end{eqnarray}
where we have dropped the constant and the higher-order terms in the magnon operators. The functions $h_{11}(q)$ and $h_{12}(q)$ are defined as
\begin{subequations}
\begin{eqnarray}
h_{11}(q) &\equiv& \frac{I}{4N_{\perp}} \left( 2 J^{z}_{q} + J^{x}_{q-2k_{F}} +  J^{x}_{q+2k_{F}} - 4J^{x}_{2k_{F}} \right) \nonumber\\
&\approx& \frac{I}{4N_{\perp}} \left( J^{x}_{q-2k_{F}} +  J^{x}_{q+2k_{F}} - 4J^{x}_{2k_{F}} \right),\\
h_{12}(q) &\equiv& \frac{I}{4N_{\perp}} \left( J^{x}_{q-2k_{F}} +  J^{x}_{q+2k_{F}} -2 J^{z}_{q} \right) \nonumber\\
&\approx& \frac{I}{4N_{\perp}} \left( J^{x}_{q-2k_{F}} +  J^{x}_{q+2k_{F}} \right),
\end{eqnarray}
\end{subequations}
where we have used the fact that the RKKY coupling is highly anisotropic $|J^{x}_{q}|,~|J^{y}_{q}| \gg |J^{z}_{q}|$. The bilinear bosonic Hamiltonian Eq.~(\ref{Eq:H_mag1}) can be diagonalized by the Bogoliubov transformation,  
\begin{eqnarray}
\left(
\begin{array}{c}
a^{\dagger}_{q} \\  a_{-q}
\end{array}
\right)
&=& 
\left( 
\begin{array}{cc}
u_{\textrm{BT}}(q) & -v_{\textrm{BT}}(q) \\
-v_{\textrm{BT}}(q) & u_{\textrm{BT}}(q) 
\end{array}
\right)
\left(
\begin{array}{c}
b^{\dagger}_{q} \\  b_{-q}
\end{array}
\right),
\label{Eq:Bogoliubov1}
\end{eqnarray}
with the coefficients,
\begin{subequations}
\label{Eq:Bogoliubov2}
\begin{eqnarray}
\left[u_{\textrm{BT}}(q)\right]^2 &=& \frac{1}{2} \left[ 1 + \frac{h_{11}(q)}{\sqrt{h_{11}^2(q) - h_{12}^2(q)} }\right], \\
\left[v_{\textrm{BT}}(q)\right]^2 &=& \frac{1}{2} \left[ -1 + \frac{h_{11}(q)}{\sqrt{h_{11}^2(q) - h_{12}^2(q)} }\right].
\end{eqnarray}
\end{subequations}
Using Eqs.~(\ref{Eq:Bogoliubov1})--(\ref{Eq:Bogoliubov2}), the magnon Hamiltonian Eq.~(\ref{Eq:H_mag1}) can be diagonalized as 
\begin{eqnarray}
\mathit{H}_{\textrm{mag}} = \frac{1}{L} \sum_{q} E_{\textrm{mag}}(q) b_{q}^{\dagger} b_{q},
\label{Eq:H_mag}
\end{eqnarray} 
with the magnon excitation energy $E_{\textrm{mag}}(q)$ given by Eq.~(\ref{Eq:omega_mag}) and shown in Fig.~\ref{Fig:magnon}. 
Since the magnon spectrum is almost dispersionless, we approximate $E_{\textrm{mag}}(q) \approx 2I|J^{x}_{2k_{F}}|/N_{\perp}$, leading to $u_{\textrm{BT}}(q) \approx 1$ and $v_{\textrm{BT}}(q) \approx 0$, and therefore $b_{q} \approx a_{q}$. The transition temperature and the temperature dependence of the order parameter can be calculated by evaluating the magnon occupation number~\cite{Braunecker:2009b,Hsu:2015}, 
\begin{equation}
 N_{\perp} \sum_{q \neq 0} \frac{1}{e^{\frac{E_{\textrm{mag}}(q)}{k_{B}T}}-1},
\end{equation}
and the results are given in Eqs.~(\ref{Eq:Bloch})--(\ref{Eq:T0}) in Sec.~\ref{SubSec:RKKY}.

\section{Transport properties of a helical Tomonaga-Luttinger liquid~\label{Appendix:transport}}

In this Appendix we sketch the calculation of the conductance and the nonlocal conductivity of the helical edge states. When the Fermi energy is away from the gap, the action is given by Eq.~(\ref{Eq:S_el}), from which we obtain the Green's function in the momentum-Matsubara frequency domain,
\begin{eqnarray}
G_{\phi\phi}(q,\omega_{n}) &=& \frac{\pi u K}{ \omega_{n}^2 + u^2q^2 },
\label{Eq:G_phi}
\end{eqnarray}
with the momentum $q$ and Matsubara frequency $\omega_{n}$. In optical measurements, the ac conductivity can be computed as in Refs.~\cite{Giamarchi:2003,Meng:2014b}, leading to 
\begin{eqnarray}
\sigma(\omega)
&=& \left. \frac{e^2}{\pi^2 \hbar} \omega_{n} G_{\phi\phi}(q=0,\omega_{n})\right|_{i\omega_{n} \rightarrow \omega+i0^{+}} \nonumber \\
&=& \frac{e^2}{\hbar} (uK) \left[ \delta \left( \omega \right) + \frac{i}{\pi} \mathcal{P} \left( \frac{1}{\omega} \right) \right],
\end{eqnarray}
as given in Eq.~(\ref{Eq:sigma_nl_omega}) in Sec.~\ref{SubSec:signature}. 

On the other hand, the nonlocal conductivity and the dc conductance of a finite-size system (in the presence of the leads) can be computed by using the Maslov-Stone approach~\cite{Maslov:1995,Ponomarenko:1995,Safi:1995}, in which the velocity and the Luttinger liquid parameter are taken to be spatially dependent, and change abruptly at the interfaces between the leads and the helical Tomonaga-Luttinger liquid. The action now takes the form,
\begin{eqnarray}
\frac{\mathit{S}_{\textrm{el}}}{\hbar} &=& \int \frac{d\tau dr}{2\pi}  \left\{ \frac{u (r)}{K(r)} \left[ \partial_{r} \phi(r,\tau) \right]^2 \right. \nonumber \\
&& \hspace{0.8in} \left. + \frac{1}{u(r)K(r) } \left[ \partial_{\tau} \phi(r,\tau) \right]^2 \right\}, 
\label{Eq:S_el_lead}
\end{eqnarray}
where the spatial dependent velocity is defined as $u(r)=v_{F}/K(r)$ with the Luttinger liquid parameter,
\begin{eqnarray}
K(r) &=& \left\{ 
\begin{array}{l}
K, ~~ \textrm{as } -L/2 \le r \le L/2, \\
K_{\textrm{L}}, ~~ \textrm{otherwise}.
\end{array}\right.
\end{eqnarray}
The charge current is related to the external electric field by Eq.~(\ref{Eq:I_sigma_E}), where the nonlocal conductivity is given by~\cite{Maslov:1995}
\begin{eqnarray}
\sigma_{\textrm{nl}}(r,r',\omega) &=& \left. \frac{e^2}{\pi^2 \hbar} \omega_{n}  G_{\textrm{nl}}(r,r',\omega_{n})\right|_{i \omega_{n} \rightarrow \omega + i 0^{+}},
\end{eqnarray}
with the nonlocal propagator $G_{\textrm{nl}}(r,r',\omega_{n})$. 
It satisfies
\begin{eqnarray}
\left\{- \partial_{r} \left[ \frac{u(r)}{K(r)} \partial_{r} \right] + \frac{\omega_{n}^2}{u(r)K(r)} \right\} G_{\textrm{nl}}(r,r',\omega_{n}) &=& \pi \delta (r-r'). \nonumber \\
\label{Eq:G_nl}
\end{eqnarray} 
and the boundary conditions,
\begin{subequations}
\begin{eqnarray}
\textrm{(i)} &~~& G_{\textrm{nl}}(r \rightarrow \pm \infty, r',\omega_{n}) = 0 ; \label{Eq:bc1} \\ 
\textrm{(ii)}&~~& G_{\textrm{nl}}(r,r',\omega_{n}) \textrm{ is continuous at } r=r',~\pm L/2 ;\label{Eq:bc2} \\
\textrm{(iii)}&~~& \frac{u(r)}{K(r)} \partial_{r} G_{\textrm{nl}}(r,r',\omega_{n}) \textrm{ is continuous at } r=\pm L/2 ;\label{Eq:bc3} \nonumber\\ \\
\textrm{(iv)}&~~& - \left.\frac{u(r)}{K(r)} \partial_{r} G_{\textrm{nl}}(r,r',\omega_{n})\right|_{r'-0^{+}}^{r'+0^{+}} = \pi. \label{Eq:bc4} 
\end{eqnarray}
\end{subequations}
We take the ansatz for the nonlocal propagator, 
\begin{eqnarray}
&& G_{\textrm{nl}}(r,r',\omega_{n}) \nonumber \\
&=& \left\{ 
\begin{array}{l} 
A e^{\frac{|\omega_{n}|r}{u_{\textrm{L}}}}, \\
B_{+} e^{\frac{|\omega_{n}|r}{u}} + B_{-} e^{-\frac{|\omega_{n}|r}{u}}, \\
C_{+} e^{\frac{|\omega_{n}|r}{u}} + C_{-} e^{-\frac{|\omega_{n}|r}{u}},\\
D e^{\frac{-|\omega_{n}|r}{u_{\textrm{L}}}},
\end{array}\right.
\begin{array}{l}
\textrm{as } r \le -L/2, \vspace{0.08in} \\
\textrm{as } -L/2 \le r \le r', \vspace{0.08in}\\
\textrm{as } r' \le r \le L/2, \vspace{0.08in}\\
\textrm{as } r \ge L/2, \vspace{0.08in}
\end{array}
\end{eqnarray}
which satisfies the condition (i). Here we define $u_{\textrm{L}} \equiv v_{F} / K_{\textrm{L}}$. The unknowns $A,$ $B_{\pm}$, $C_{\pm}$, and $D$ are functions of $r'$ and $\omega_{n}$, and can be solved for by applying the boundary conditions (ii)--(iv)~\cite{Maslov:1995,Ponomarenko:1995,Safi:1995,Meng:2014b,Muller:2017}. 
The solutions for these unknowns are
\begin{widetext}
\begin{subequations}
\begin{eqnarray}
A &=& \frac{\pi K}{2 |\omega_{n}|} e^{\frac{|\omega_{n}|L}{2u_{\textrm{L}}}} 
\left[ \frac{\cosh \left( \frac{|\omega_{n}|r'}{u} \right) }{ \sinh 
\left( \frac{|\omega_{n}|L}{2u} \right)
+\frac{K}{K_{\textrm{L}}} \cosh \left( \frac{|\omega_{n}|L}{2u} \right) } 
- \frac{\sinh \left( \frac{|\omega_{n}|r'}{u} \right)  }{ \cosh \left( \frac{|\omega_{n}|L}{2u} \right) +\frac{K}{K_{\textrm{L}}} \sinh \left( \frac{|\omega_{n}|L}{2u} \right) }\right], \\
B_{\pm} &=& \frac{\pi K}{4 |\omega_{n}|} e^{\pm \frac{|\omega_{n}|L}{2u}} 
\left( 1 \pm \frac{K}{K_{\textrm{L}}}\right) \left[ \frac{\cosh \left( \frac{|\omega_{n}|r'}{u} \right) }{ \sinh 
\left( \frac{|\omega_{n}|L}{2u} \right)
+\frac{K}{K_{\textrm{L}}} \cosh \left( \frac{|\omega_{n}|L}{2u} \right) } 
- \frac{\sinh \left( \frac{|\omega_{n}|r'}{u} \right)  }{ \cosh \left( \frac{|\omega_{n}|L}{2u} \right) +\frac{K}{K_{\textrm{L}}} \sinh \left( \frac{|\omega_{n}|L}{2u} \right) }\right], \nonumber\\ \\
C_{\pm} &=& \frac{\pi K}{4 |\omega_{n}|} e^{\mp \frac{|\omega_{n}|L}{2u}} 
\left( 1 \mp \frac{K}{K_{\textrm{L}}}\right) \left[ \frac{\cosh \left( \frac{|\omega_{n}|r'}{u} \right) }{ \sinh 
\left( \frac{|\omega_{n}|L}{2u} \right)
+\frac{K}{K_{\textrm{L}}} \cosh \left( \frac{|\omega_{n}|L}{2u} \right) } 
+ \frac{\sinh \left( \frac{|\omega_{n}|r'}{u} \right)  }{ \cosh \left( \frac{|\omega_{n}|L}{2u} \right) +\frac{K}{K_{\textrm{L}}} \sinh \left( \frac{|\omega_{n}|L}{2u} \right) }\right], \nonumber\\ \\
D &=& \frac{\pi K}{2 |\omega_{n}|} e^{\frac{|\omega_{n}|L}{2u_{\textrm{L}}}} 
\left[ \frac{\cosh \left( \frac{|\omega_{n}|r'}{u} \right) }{ \sinh 
\left( \frac{|\omega_{n}|L}{2u} \right)
+\frac{K}{K_{\textrm{L}}} \cosh \left( \frac{|\omega_{n}|L}{2u} \right) } 
+ \frac{\sinh \left( \frac{|\omega_{n}|r'}{u} \right)  }{ \cosh \left( \frac{|\omega_{n}|L}{2u} \right) +\frac{K}{K_{\textrm{L}}} \sinh \left( \frac{|\omega_{n}|L}{2u} \right) }\right].
\end{eqnarray}
\end{subequations}
\end{widetext} 
For the dc signals the $r$ and $r'$ dependence in the nonlocal conductivity will eventually vanish, allowing us to focus on the origin [$(r,r')=(0,0)$]. We then get the propagator,
\begin{eqnarray}
 G_{\textrm{nl}}(0,0,\omega_{n}) &=& \frac{\pi K}{2 |\omega_{n}|}
\frac{ 1+ \frac{K}{K_{\textrm{L}}} \tanh \left( \frac{|\omega_{n}|L}{2u} \right) }{ \frac{K}{K_{\textrm{L}}} + \tanh \left( \frac{|\omega_{n}|L}{2u} \right)},
\end{eqnarray}
and the nonlocal conductivity $\sigma_{\textrm{nl}}(0,0,\omega)$, as given in Eq.~(\ref{Eq:sigma_nl}). The dc conductance is then $G_{\textrm{dc}} = \lim_{\omega \rightarrow 0} \textrm{Re}[\sigma_{\textrm{nl}}(0,0,\omega)] = K_{\textrm{L}} e^2/h $, as given in Eq.~(\ref{Eq:G_lead}).

When the Fermi energy is quickly tuned into the gap, the action acquires an RG-relevant sine-Gordon term,
\begin{eqnarray}
\delta S_{\textrm{m}} &=& 
\frac{B_{\textrm{Ov}}}{2\pi a} \int_{0}^{\beta \hbar} d\tau \int dr \;  \cos \left[ 2 \phi (r,\tau) \right],
\label{Eq:S_SG}
\end{eqnarray}
leading to the propagator, 
\begin{eqnarray}
G_{\phi\phi}(q,\omega_{n}) &=& \frac{\pi u K}{ \omega_{n}^2 + u^2q^2 + \Delta_{\textrm{m}}^2/\hbar^2}.
\label{Eq:G_phi_gap}
\end{eqnarray}
The dc conductance and the ac conductivity can then be computed following the same procedure, and the results are given in Sec.~\ref{SubSec:signature}.

\section{Schrieffer-Wolff transformation~\label{Appendix:SW}}

In this Appendix we perform the Schrieffer-Wolff transformation to obtain the effective Hamiltonian for the Overhauser-field-assisted backscattering on impurities. In the absence of the Overhauser field, (nonmagnetic) impurities cannot cause the spin-flip backscattering, so the helical edge states cannot be localized by the impurities. The Overhauser field, however, acts on electron spins as a spatially rotating Zeeman field, which breaks the time-reversal symmetry. Assuming that the nuclear spin order is given by $\left\langle {\bf \tilde{I}}(r) \right\rangle_{-}$ [see Eq.~(\ref{Eq:order})], the Overhauser field then causes a mixing of the $R_{\downarrow}(q)$ and $L_{\uparrow}(q+2k_{F})$ particles, inducing a gap $\Delta_{\textrm{m}}$ below the Fermi surface [panel (b) of Fig.~\ref{Fig:helicity}]. Whereas Eq.~(\ref{Eq:H_Ov}) itself does not lead to any backscattering at the Fermi surface, here we show that a second-order spin-flip backscattering at the Fermi surface can still arise as a combination of the Overhauser field and the impurities, as sketched in Fig.~\ref{Fig:backscattering}. Here `second-order' means that the effective backscattering potential is determined by the product of the Overhauser field and the impurity potential.

To proceed, we consider the total Hamiltonian, which consists of two parts, $\mathit{H}_{\textrm{tot}} = \mathit{H}_{\textrm{el}} + \delta\mathit{V}$, with the perturbation $\delta\mathit{V}=\mathit{H}_{\textrm{imp}} + \mathit{H}_{\textrm{Ov}} $. 
For convenience, we use the fermionic expression for the electron part, $\mathit{H}_{\textrm{el}}=\mathit{H}_{0}+\mathit{H}_{2}+\mathit{H}_{4}$, with the three terms corresponding to the kinetic energy, $g_{2}$, and $g_{4}$ processes, respectively. To be explicit, we have
\begin{subequations}
\label{Eq:H_024}
\begin{align}
&\mathit{H}_{0} = -i \hbar v_{F} \int dr \left[
R_{\downarrow}^{\dagger}(r) \partial_{r} R_{\downarrow}(r) - L_{\uparrow}^{\dagger}(r) \partial_{r} L_{\uparrow}(r) \right], \\ 
&\mathit{H}_{2} = g_{2} \int dr \; R_{\downarrow}^{\dagger}(r)  R_{\downarrow}(r)  L_{\uparrow}^{\dagger}(r)  L_{\uparrow}(r),  \\
&\mathit{H}_{4} = \frac{g_{4}}{2} \int dr \left[
\left(R_{\downarrow}^{\dagger}(r)  R_{\downarrow}(r) \right)^2 + \left( L_{\uparrow}^{\dagger}(r)  L_{\uparrow}(r) \right)^2 \right]. 
\end{align}
\end{subequations}
The impurity Hamiltonian $\mathit{H}_{\textrm{imp}}$ is given by Eq.~(\ref{Eq:H_imp}), and the Overhauser field felt by the electrons is described by
\begin{align}
\mathit{H}_{\textrm{Ov}} \equiv \langle \mathit{H}_{\textrm{hf}} \rangle_{\textrm{gs}} &=& \frac{B_{\textrm{Ov}}}{2} \sum_{j} \; e^{2ik_{F}r_{j}} L_{\uparrow}^{\dagger}(r_{j}) R_{\downarrow}(r_{j}) + \textrm{H.c.}, 
\label{Eq:H_Ov2}
\end{align}
which is the fermionic form of $\langle \mathit{H}_{\textrm{hf}} \rangle_{\textrm{gs}} $ in Eq.~(\ref{Eq:H_Ov}).

We then perform the canonical Schrieffer-Wolff transformation~\cite{Schrieffer:1966,Bravyi:2011}  such that
\begin{eqnarray}
\mathit{H}_{\textrm{SW}} &\equiv& e^{S_{\textrm{SW}}} \mathit{H}_{\textrm{tot}} e^{-S_{\textrm{SW}}} \nonumber\\
&=& \mathit{H}_{\textrm{el}} + \delta\mathit{V} + [S_{\textrm{SW}},\mathit{H}_{\textrm{el}} + \delta\mathit{V}] \nonumber\\
&& + \frac{1}{2} [S_{\textrm{SW}},[S_{\textrm{SW}}, \mathit{H}_{\textrm{el}} + \delta\mathit{V}]] + \cdots,
\end{eqnarray}
where we keep terms up to the second order in $\delta\mathit{V}$ and $S_{\textrm{SW}}$, and choose $[S_{\textrm{SW}}, \mathit{H}_{\textrm{el}}] + \delta\mathit{V}=0$ to eliminate the first-order term in $\delta\mathit{V}$. This gives $S_{\textrm{SW}} = \mathcal{L}_{\textrm{el}}^{-1} \delta\mathit{V} $ with the Liouvillian superoperator $\mathcal{L}_{\textrm{el}} \mathit{O} \equiv [\mathit{H}_{\textrm{el}}, \mathit{O}]$~\cite{Simon:2008}. 
Using the integral representation, $ \mathcal{L}_{\textrm{el}}^{-1} = -i \int_{0}^{\infty} dt \; \left.e^{-\eta t + i t \mathcal{L}_{\textrm{el}}}\right|_{\eta \rightarrow 0}$, we arrive at the effective Hamiltonian $\mathit{H}_{\textrm{SW}} \approx \mathit{H}_{\textrm{el}} + \mathit{H}_{\textrm{hx}}$, where the backscattering term is given by 
\begin{align}
\mathit{H}_{\textrm{hx}} \equiv&\frac{1}{2} [S_{\textrm{SW}},\delta\mathit{V}] 
= \frac{-i}{2} \int_{0}^{\infty} dt \; e^{-\eta t}
[\delta\tilde{\mathit{V}}(t),\delta\mathit{V}]_{\eta \rightarrow 0}, 
\end{align}
with the tilde defining an operator in the interaction picture, $\delta\tilde{\mathit{V}}(t) \equiv \tilde{\mathit{H}}_{\textrm{imp}}(t) + \tilde{\mathit{H}}_{\textrm{Ov}} (t)$. The commutators can be computed straightforwardly, and the results can be simplified by approximating $g_{2} \approx g_{4}$. This approximation can be understood through the form of Eq.~(\ref{Eq:H_024}). Namely, for the Hamiltonian involving only the density-density interaction and therefore insensitive to the spins and the velocities of the electrons, the $g_2$ and $g_4$ processes are indistinguishable.  

After performing the integral over time, we finally arrive at
\begin{align}
\mathit{H}_{\textrm{hx}}
&\approx& \frac{1}{L^2} \sum_{q,q'} \; V_{\textrm{hx}}(q') R_{\downarrow}^{\dagger}(q+q'+2k_{F}) L_{\uparrow}(q) + \textrm{H.c.}, 
\end{align}
with the effective coupling for the second-order backscattering process $V_{\textrm{hx}} (q) \equiv B_{\textrm{Ov}} V_{\textrm{imp}} (4k_{F}+q)/ (8 \hbar v_{F} k_{F}) $. 
After performing inverse Fourier transform and bosonizing $\mathit{H}_{\textrm{hx}}$, the $q' \neq 0$ components give oscillating integrand, and therefore vanish upon integration. Consequently, the effective backscattering potential is given by the product of $B_{\textrm{Ov}}/2$ [the strength of $\mathit{H}_{\textrm{Ov}}$ in Eq.~(\ref{Eq:H_Ov2})], and $V_{\textrm{imp}} (4k_{F})$ [the $4k_{F}$ component of the random potential in Eq.~(\ref{Eq:H_imp})], divided by the energy difference between the initial and the intermediate states, $4\hbar v_{F} k_{F}$, as expected from Fig.~\ref{Fig:backscattering}. Finally, utilizing the replica method~\cite{Giamarchi:2003}, we can average the random potential $V_{\textrm{imp}}$ in $\mathit{H}_{\textrm{hx}}$, leading to the effective backscattering action Eq.~(\ref{Eq:S_hx-bs}).

\section{Magnon-mediated backscattering~\label{Appendix:mag-bs}}

In this Appendix we provide the derivation of the effective action for the magnon-mediated backscattering process. In the bosonized form and the continuum limit, the electron-magnon interaction, described by Eq.~\eqref{Eq:H_e-mag_FT}, can be written as 
\begin{eqnarray}
\mathit{H}_{\textrm{e-mag}} 
&=& 
\int \frac{dr }{2\pi a} \; \left[ g_{\textrm{e-mag}} \varphi (r) e^{-2i\phi (r)} + \textrm{H.c.} \right],
\end{eqnarray}
where we introduce the effective electron-magnon coupling,
\begin{equation}
g_{\textrm{e-mag}} \equiv -i \frac{A_{0}}{2} \sqrt{\frac{\omega_{\textrm{mag}} I}{\hbar N_{\perp}}},
\end{equation}
and the bosonic field,
\begin{equation}
\varphi (r) \equiv \sqrt{ \frac{\hbar}{2 \omega_{\textrm{mag}}} } \frac{1}{L} \sum_{ q} b_{q} e^{i q r} + \textrm{H.c.}.
\end{equation}
Here $\varphi$ is analogous to the displacement field in the electron-phonon problem~\cite{Voit:1987,Voit:1995}. 
With these definitions, the magnon Hamiltonian [see Eq.~(\ref{Eq:H_mag})] can be written, up to a constant term, as
\begin{align}
\mathit{H}_{\textrm{mag}}  
=\frac{1}{2} \int dr \; \left[ \Pi^2 (r) + \omega_{\textrm{mag}}^2 \varphi^2 (r) \right],
\end{align}
with $\Pi$ being the canonically conjugate momentum to $\varphi$. In the above we have used the fact that both $\varphi$ and $\Pi$ are Hermitian.
Therefore, the terms involving the magnons, $\mathit{H}_{\textrm{mag}}+\mathit{H}_{\textrm{e-mag}} $, lead to the contribution to the imaginary-time action $\delta \mathit{S}_{\textrm{mag}} \equiv \delta \mathit{S}_{\textrm{mag}}^{(0)} + \delta \mathit{S}_{\textrm{mag}}^{(1)}$, where 
\begin{eqnarray}
\frac{\delta \mathit{S}_{\textrm{mag}}^{(0)} }{\hbar}&=& 
\frac{1}{2} \int dr d\tau\; \left[ -2i \Pi(r,\tau) \partial_{\tau}\varphi (r,\tau) + \Pi^2 (r,\tau) \right. \nonumber \\
&& \hspace{0.6in} \left. + \omega_{\textrm{mag}}^2 \varphi^2 (r,\tau) \right], \\
\frac{\delta \mathit{S}_{\textrm{mag}}^{(1)} }{\hbar}&=& \int \frac{dr d\tau }{2\pi a\hbar} \; \left[ g_{\textrm{e-mag}} \varphi (r,\tau) e^{-2i\phi (r,\tau)} + \textrm{H.c.}\right]. \end{eqnarray}
We first integrate out the $\Pi$ field in the term $\delta \mathit{S}_{\textrm{mag}}^{(0)}$. In the momentum space and Matsubara frequency domain, it is given by
\begin{eqnarray}
\frac{\delta \mathit{S}_{\textrm{mag}}^{(0)} }{\hbar}
&=& 
\frac{1}{2\beta L} \sum_{q,\omega_{n}} \; 
\left[ \delta G_{\textrm{mag}} (q,\omega_{n}) \right]^{-1} \left| \varphi (q,\omega_{n}) \right|^2 ,
\end{eqnarray}
where the magnon propagator is defined as
\begin{eqnarray}
\delta G_{\textrm{mag}} (q,\omega_{n}) &=& - \frac{1}{(i\omega_{n})^2 - \omega_{\textrm{mag}}^2},
\end{eqnarray}
which is independent of the momentum, as we are considering the magnons with dispersionless energy band. Finally, by integrating out the remaining $\varphi$ field in the action $\delta \mathit{S}_{\textrm{mag}}$, we get 
\begin{eqnarray}
\frac{\delta \mathit{S}_{\textrm{mag}} }{\hbar} 
&=& -\frac{\left| g_{\textrm{e-mag}} \right|^2 }{(2\pi)^2 a\hbar} 
\int dr dr' d\tau d\tau' \;  \delta G_{\textrm{mag}} (r-r',\tau - \tau') \nonumber\\
&& \hspace{0.7in} \times \left[ e^{-2i[\phi(r,\tau) - \phi(r',\tau') ]} + \textrm{H.c.} \right],
\end{eqnarray}
with the magnon propagator,
\begin{eqnarray}
\delta G_{\textrm{mag}} (r,\tau) &=& \frac{1}{\beta L} \sum_{q,\omega_{n}} \; \delta G_{\textrm{mag}} (q,\omega_{n}) e^{i(qr-\omega_{n}\tau)} .
\end{eqnarray}
The summation over the momentum and the Matsubara frequency can be done straightforwardly~\cite{Bruus:2004}, which leads to
\begin{align}
&\delta G_{\textrm{mag}} (r,\tau) =  \frac{1}{2 \omega_{\textrm{mag}}}
\delta(r)  \nonumber\\
&\hspace{20pt} \times \left[ e^{-\omega_{\textrm{mag}}|\tau|} + 2 n_{B}(\hbar \omega_{\textrm{mag}}) \cosh \left( \omega_{\textrm{mag}} \tau \right) \right].
\label{Eq:G_mag_T}
\end{align}
The resulting effective magnon-mediated backscattering action is then given in Eq.~(\ref{Eq:S_mag-bs}) in Sec.~\ref{SubSec:mag-bs}. 

Since the magnon energy $\hbar \omega_{\textrm{mag}}$, see Eq.~(\ref{Eq:omega_mag_T}), depends on the temperature, the behavior and the validity of Eq.~(\ref{Eq:S_mag-bs}) also depend on the temperature. First, in the low-temperature regime where $k_{B}T \ll \hbar \omega_{\textrm{mag}}$, we may approximate $n_{B}(\hbar \omega_{\textrm{mag}}) \approx 0$ to obtain the effective action dominated by magnon emission, from which we derive Eq.~(\ref{Eq:R_mag-em}). Second, in the $T \apprle T_{0}$ regime, the magnon energy is so low that the procedure of integrating out the magnon fields is no longer valid. In this case we compute Eq.~(\ref{Eq:R_mag-abs}) using the resistance in the disordered phase, $R_{\textrm{hf}}$, which can be considered as the resistance due to the ordered nuclear spins, but with zero-energy magnons.


%

\end{document}